%% file: paper.tex
\documentclass[10pt,journal,compsoc]{IEEEtran}

\input{setup}

\begin{document}

\title{Studying Duplicate Logging Statements and Their Relationships with Code Clones}
%\title{Studying Duplicate Logging Code Smells and Their Relationships with Code Clones}

\author{Zhenhao~Li,~\IEEEmembership{Student Member,~IEEE,}
        Tse-Hsun~(Peter)~Chen,~\IEEEmembership{Member,~IEEE,}
        Jinqiu~Yang,~\IEEEmembership{Member,~IEEE,}
        and~Weiyi~Shang,~\IEEEmembership{Member,~IEEE}
\IEEEcompsocitemizethanks{\IEEEcompsocthanksitem Z. Li, T. Chen J. Yang and W. Shang are with the Department of Computer Science and Software Engineering, Concordia University, Montreal, Quebec, Canada.\protect\\
E-mail: {l\_zhenha,peterc,jinqiuy,shang}@encs.concordia.ca 
}% <-this % stops an unwanted space
%\thanks{Manuscript received xxx, 2018; revised xxx, 2018.}
}

\IEEEtitleabstractindextext{%
\begin{abstract}
\input{texfiles/abstract}

\end{abstract}

}

\maketitle

\IEEEdisplaynontitleabstractindextext

\IEEEpeerreviewmaketitle

% ------------Main sections------------
\IEEEraisesectionheading{\section{Introduction}\label{sec:intro}}

\input texfiles/intro

\input texfiles/prestudy
\input texfiles/pattern

\input texfiles/detection

\input texfiles/results

\input texfiles/rq4

\input texfiles/rq5
\input texfiles/threats
\input texfiles/related

\input texfiles/conclusion
\bibliographystyle{IEEEtran}
\footnotesize
\bibliography{paper}

% ------------bio info------------
\begin{IEEEbiography}[{\includegraphics[width=1in,height=1.25in,clip,keepaspectratio]{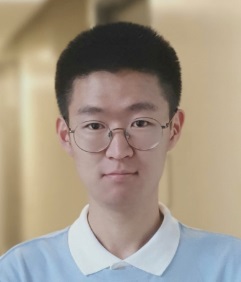}}]{Zhenhao Li}
Zhenhao Li is a Ph.D. student at the Department of Computer Science and Software Engineering at Concordia University, Montreal, Canada. He obtained his M.ASc degree from Concordia University and B.Eng. from Harbin Institute of Technology. His work has been published at renowned venues such as ICSE and ASE.
His research interests include software log analysis, improving logging practices, program analysis, and mining software repositories. More information at: https://ginolzh.github.io/.
\end{IEEEbiography}

\begin{IEEEbiography}[{\includegraphics[width=1in,height=1.25in,clip,keepaspectratio]{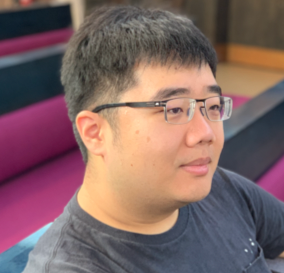}}]{Tse-Hsun (Peter) Chen}
Tse-Hsun (Peter) Chen is an Assistant Professor in the Department of Computer Science and Software Engineering at Concordia University, Montreal, Canada. He leads the Software PErformance, Analysis, and Reliability (SPEAR) Lab, which focuses on conducting research on performance engineering, program analysis, log analysis, production debugging, and mining software repositories. His work has been published in flagship conferences and journals such as ICSE, FSE, TSE, EMSE, and MSR. He serves regularly as a program committee member of international conferences in the field of software engineering, such as ASE, ICSME, SANER, and ICPC, and he is a regular reviewer for software engineering journals such as JSS, EMSE, and TSE. Dr. Chen obtained his BSc from the University of British Columbia, and MSc and PhD from Queen's University. Besides his academic career, Dr. Chen also worked as a software performance engineer at BlackBerry for over four years. Early tools developed by Dr. Chen were integrated into industrial practice for ensuring the quality of large-scale enterprise systems. More information at: https://petertsehsun.github.io/.
\end{IEEEbiography}

\begin{IEEEbiography}[{\includegraphics[width=1in,height=1.25in,clip,keepaspectratio]{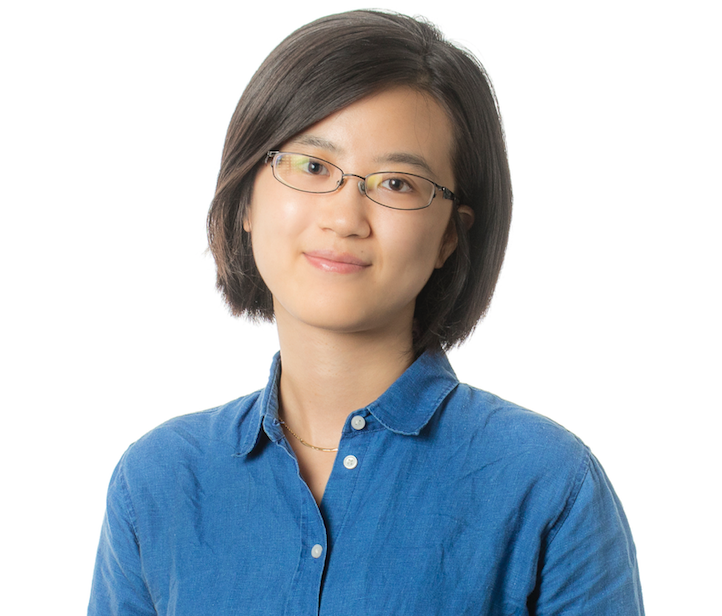}}]{Jinqiu Yang}
        Jinqiu Yang is an Assistant Professor in the Department of Computer Science and Software Engineering at Concordia University, Montreal, Canada. Her research interests include automated program repair, software testing, software text analytics, and mining software repositories. Her work has been published flagship conferences and journals such as ICSE, FSE, EMSE. She serves regularly as a program committee member of international conferences in Software Engineering, such as ASE, ICSE, ICSME and SANER. She is a regular reviewer for Software Engineering journals such as EMSE and JSS. Dr. Yang obtained her BEng from Nanjing University, and MSc and PhD from University of Waterloo. More information at: https://jinqiuyang.github.io/.
\end{IEEEbiography}

\begin{IEEEbiography}[{\includegraphics[width=1in,height=1.25in,clip,keepaspectratio]{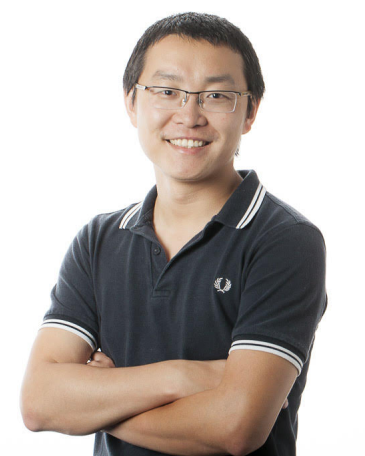}}]{Weiyi Shang}
Weiyi Shang is an Assistant Professor and Concordia University Research Chair in Ultra-large-scale Systems at the Department of Computer Science and Software Engineering at Concordia University, Montreal.
He has received his Ph.D. and M.Sc. degrees from Queens University (Canada) and he obtained B.Eng.
from Harbin Institute of Technology. His research interests include big data software engineering, software
engineering for ultra-largescale systems, software log mining, empirical software engineering, and software
performance engineering. His work has been published at premier venues such as ICSE, FSE, ASE, ICSME,
MSR and WCRE, as well as in major journals such as TSE, EMSE, JSS, JSEP and SCP. His work has won
premium awards, such as SIGSOFT Distinguished paper award at ICSE 2013 and best paper award at WCRE
2011. His industrial experience includes helping improve the quality and performance of ultra-large-scale
systems in BlackBerry. Early tools and techniques developed by him are already integrated into products
used by millions of users worldwide. Contact him at shang@encs.concordia.ca; \url{https://users.encs.concordia.ca/~shang}.
\end{IEEEbiography}

\clearpage

\appendices

\section{Precision of NiCad on Detecting Duplicate Logging Statements that Reside in Cloned Code}
\label{sec:appendix2}
We rely on NiCad for automated clone detection. To examine the false positives of NiCad, we then manually verify a randomly sampled set of duplicate logging statements (281 sets in total, with 95\% confidence level and 5\% confidence interval) that are classified as clones by NiCad. For each set of the sampled duplicate logging statements, we manually go through the logging statements and their surrounding code to verify whether they are clones or not. Overall, we find that 272 out of the 281 sampled sets (96.8\%) are clones, which is similar to the performance of NiCad that is reported in prior studies. For the 9 false positives, 3 of them are duplicate logging statements located in different branches of a nested method (i.e., developers define a method within a method). In such cases, NiCad would analyze the code block twice. For example, in ElasticSearch~\footnote{\url{https://github.com/elastic/elasticsearch/blob/70b8d7bc64f165735502de9d8c5fa673fa21e02b/server/src/main/java/org/elasticsearch/cluster/InternalClusterInfoService.java}}, two duplicate logging statements with the same static text message {\em ``Failed to execute NodeStatsAction for ClusterInfoUpdateJob''} are located in different branches of the same nested method {\em onFailure(Exception e)}. However, since the method {\em onFailure(Exception e)} is defined in the method {\em refresh()}, NiCad would analyze the same code block twice and detect them as clones. For the remaining 6 out of 9 false positives, we could not identify the reasons that they are classified as clones, since the code snippets look neither structurally nor semantically similar.

\end{document}

%% file: setup.tex
\usepackage{listings}
\usepackage{tcolorbox}

\usepackage{verbatim}
\usepackage{stfloats}
\usepackage{textcomp}
\usepackage{booktabs, tabularx}
\usepackage{rotating}
\usepackage{soul}
\usepackage{xcolor}
\usepackage{cite}

\usepackage{fancybox}
 \usepackage{multirow}

\usepackage{url}
\usepackage{hyperref}

\usepackage[caption=false,font=footnotesize]{subfig}
\usepackage{lipsum}
\usepackage{graphicx}
\usepackage{balance}
\usepackage{footnote}

\usepackage{array}

\usepackage{adjustbox}
\usepackage{threeparttable}

\newcommand{\phead}[1]{\vspace{1mm} \noindent {\bf #1}}

\newcommand{\peter}[1]{\textcolor{red}{{\it [Peter says: #1]}}}

\newcommand{\jinqiu}[1]{\textcolor{red}{{\it [Jinqiu says: #1]}}}

\newcommand{\toolS}{DLFinder\space}
\newcommand{\tool}{DLFinder}
\newcommand{\rqboxc}[1]{\begin{tcolorbox}[left=4pt,right=4pt,top=4pt,bottom=4pt,colback=gray!5,colframe=gray!40!black,before skip=2pt,after skip=2pt]#1\end{tcolorbox}}

\definecolor{dkgreen}{rgb}{0,0.6,0}
\definecolor{gray}{rgb}{0.5,0.5,0.5}
\definecolor{mauve}{rgb}{0.58,0,0.82}

\usepackage{array}
\newcolumntype{$}{>{\global\let\currentrowstyle\relax}}
\newcolumntype{^}{>{\currentrowstyle}}

%% file: texfiles/abstract.tex
%Developers use software logging statements to record system execution.
Developers rely on software logs for a variety of tasks, such as debugging, testing, program comprehension, verification, and performance analysis. Despite the importance of logs, prior studies show that there is no industrial standard on how to write logging statements. %Recent research on logs often only considers the appropriateness of a log as an individual item (e.g., one single logging statement); while logs are typically analyzed in sequences or clusters. %Similar to other part of the source code, there may be certain code smells that are associated with logging code. Even though there is an established line of research on code smells such as duplicate code, these studies do not help developers refactor possible code smells in logging code.
In this paper, we focus on studying duplicate logging statements, which are logging statements that have the same static text message. Such duplications in the text message are potential indications of logging code smells, which may affect developers' understanding of the dynamic view of the system. We manually studied over 4K duplicate logging statements and their surrounding code in five large-scale open source systems: Hadoop, CloudStack, Elasticsearch, Cassandra, and Flink. We uncovered five patterns of duplicate logging code smells. For each instance of the duplicate logging code smell, we further manually identify the potentially problematic (i.e., require fixes) and justifiable (i.e., do not require fixes) cases. Then, we contact developers to verify our manual study result. %Developers confirmed our findings in general while considered that some of the potentially problematic instances are related to technical debt. 
We integrated our manual study result and developers' feedback into our automated static analysis tool, \tool, which automatically detects problematic duplicate logging code smells. We evaluated \toolS on the five manually studied systems and three additional systems: Camel, Kafka and Wicket. In total, combining the results of \toolS and our manual analysis, we reported 91 problematic duplicate logging code smell instances to developers and all of them have been fixed. We further study the relationship between duplicate logging statements, including the problematic instances of duplicate logging code smells, and code clones. We find that 83\% of the duplicate logging code smell instances reside in cloned code, but 17\% of them reside in micro-clones that are difficult to detect using automated clone detection tools. We also find that more than half of the duplicate logging statements reside in cloned code snippets, and a large portion of them reside in very short code blocks which may not be effectively detected by existing code clone detection tools. Our study shows that, in addition to general source code that implements the business logic, code clones may also result in bad logging practices that could increase maintenance difficulties. %We find that nearly half of the duplicate logging statements reside in cloned code snippets, and a large portion of them reside in very short code blocks which may not be effectively detected by existing code clone detection tools. We also find that most of the problematic instances of duplicate logging code smells reside in cloned code snippets, which indicates that code clones may lead to bad logging practices that could increase maintenance difficulties. 

%% file: texfiles/intro.tex
\IEEEPARstart{S}{oftware} logs are widely used in software systems to record system execution behaviors. Developers use the generated logs to assist in various tasks, such as debugging~\cite{Yuan:2011:ISD:1950365.1950369, Yuan:2010:SED:1736020.1736038, Fu:2014:DLE:2591062.2591175,gino_DS, ARC_log_bug_report, cyfICSE2019,hpj_log_analysis_TDSC18,hpj_ISSRE16}, testing~\cite{Chen:2017:ALT:3103112.3103144, jacktool, jackase2018,JF_ASE19}, program comprehension~\cite{Hassan:2008:ICS:1368088.1379445, Shang:2014:ULL:2705615.2706065,joy_mobile_log}, system verification~\cite{Busany:2016:BLA:2884781.2884805, DBLP:journals/jacic/BarringerGHS10}, and performance analysis~\cite{Chen:2016:CHD:2950290.2950303, kundi_icpe_2018, zs_performance_test,kundi_log_compression}. A logging statement (i.e., code that generates a log) contains a static message, to-be-recorded variables, and log verbosity level. For example, in the logging statement: {\em logger.error(``Interrupted while waiting for fencing command: '' + cmd)}, the static text message is {\em ``Interrupted while waiting for fencing command:''}, and the dynamic message is from the variable {\em cmd}, which records the command that is being executed. The logging statement is at the {\em error} level, which is the level for recording failed operations~\cite{log4j}.

Even though developers have been analyzing logs for decades~\cite{Kabinna:2016:LLM:2901739.2901769}, there exists no industrial standard on how to write logging statements~\cite{Fu:2014:DLE:2591062.2591175, 7202961}. Prior studies often focus on recommending where logging statements should be added into the code (i.e., {\em where-to-log})~\cite{Zhu:2015:LLH:2818754.2818807, wheretologASE, wheretologSRC, Zhao:2017:LFA:3132747.3132778}, and what information should be added in logging statements (i.e., {\em what-to-log})~\cite{Shang:2014:ULL:2705615.2706065, Yuan:2011:ISD:1950365.1950369, aseLog2018,loglevel_ICSE21}. A few recent studies~\cite{log_pattern_ICSE2017, mehran_emse_2018} aim to detect potential problems in logging statements. However, these studies often only consider the appropriateness of one single logging statement; while logs are typically analyzed in sequences or clusters~\cite{Yuan:2011:ISD:1950365.1950369, Chen:2016:CHD:2950290.2950303}. In other words, we consider that the appropriateness of a log is also influenced by other logs that are generated in system execution.

In particular, an intuitive case of such influence is duplicate logs, i.e., multiple logs that have the same text message. Even though each log itself may be impeccable, duplicate logs, in some occasions, may affect developers' understanding of the dynamic view of the system. For example, as shown in Figure~\ref{fig:M_example_intro}, there are two logging statements in two different {\em catch} blocks, which are associated with the same {\em try} block. These two logging statements have the same static text message and do not include any other error-diagnostic information. Thus, developers cannot easily distinguish what is the occurred exception when analyzing the produced logs.
%we find that developers copied and pasted a log statement from {\em removeGroup} to {\em updateGroup} (i.e., causes duplicate logs), but the copied log statement in {\em updateGroup} still contains the static text {\em ``remove group''}, which should be updated in the new context. %Although the two methods have syntactic similarities, the semantic are different (i.e., update v.s. remove).
Since developers rely on logs for debugging and program comprehension~\cite{Shang:2014:ULL:2705615.2706065}, such duplicate logging statements may negatively affect developers' activities in maintenance and quality assurance. %On the other hand, developers may also intentionally inject such duplicate logs to accomplish tasks that are not easily supported by the current logging libraries.

 \begin{figure}
 \centering
 %\begin{lstlisting}
%public boolean updateGroup(final CloudianGroup group) {
 %   ...
  %  } catch (final IOException e) {
   %     LOG.error("Failed to remove group due to:", e);
    %    checkResponseTimeOut(e);
    %}
    %...
%}
 %\end{lstlisting}
%\includegraphics[width=0.9\linewidth]{figures/M_example_intro.pdf}

\begin{lstlisting}
...
} catch (AlreadyClosedException closedException) {
       s_logger.warn("Connection to AMQP service is lost.");
} catch (ConnectException connectException) {
       s_logger.warn("Connection to AMQP service is lost.");
}
...
\end{lstlisting}
\vspace{-0.3cm}
 \caption{An example of duplicate logging code smell that we detected in CloudStack. The duplicate logging statements in the two {\em catch} blocks contain insufficient information (e.g., no exception type or stack trace) to distinguish the occurred exception.}
 \vspace{-0.3cm}
 \label{fig:M_example_intro}
 \end{figure}

 \begin{figure}
 \centering
\includegraphics[width=\columnwidth]{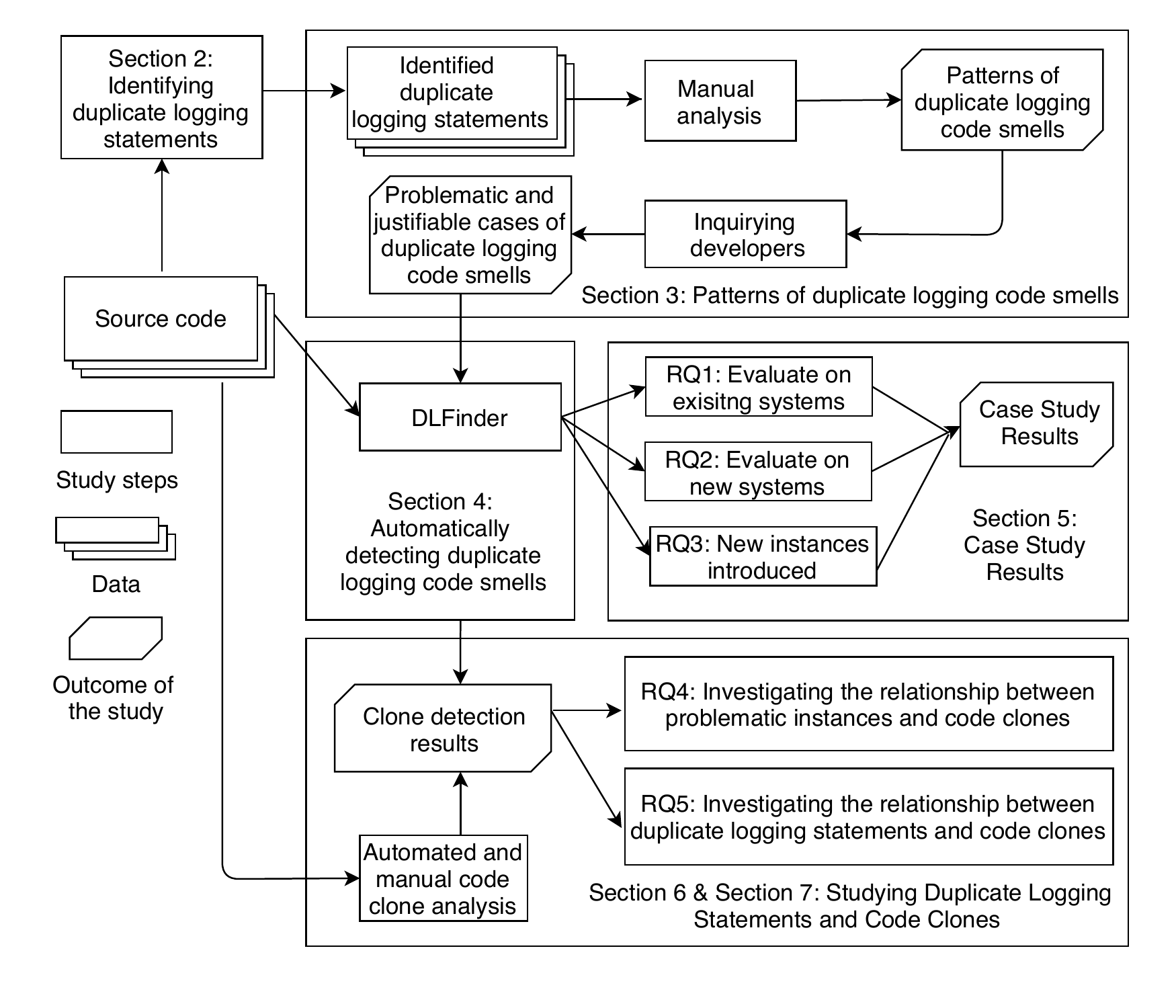}
\vspace{-0.6cm}
 \caption{The overall process of our study.}
 \vspace{-0.6cm}
 \label{fig:overall}
 \end{figure}

%In order to understand the phenomenon of log duplication in depth,
%\ian{inconsistent between present and past tense, stick to one.} \zhenhao{converted to past tense}
To help developers improve logging practices, in this paper, we focus on studying duplicate logging statements in the source code. We conducted a manual study on five large-scale open source systems, namely Hadoop, CloudStack, ElasticSearch, Cassandra and Flink. We first used static analysis to identify all duplicate logging statements, which are defined as two or more logging statements that have the same static text message. We then manually study all the (over 4K) identified duplicate logging statements and uncovered five patterns of {\em duplicate logging code smells}. We follow prior code smell studies~\cite{budgen2003software, fowler1999refactoring}, and consider duplicate logging code smell as a ``surface indication that usually corresponds to a deeper problem in the system''. Thus, {\em not} all of the duplicate logging code smell are problematic and require fixes (i.e., {\em problematic duplicate logging code smells}). In particular, context (e.g., surrounding code and usage scenario of logging) may play an important role in identifying fixing opportunities. We further categorized duplicate logging code smells into potentially {\em problematic} or {\em justifiable} cases. In addition to our manual analysis, we sought confirmation from developers on the manual analysis result. For some of the potentially problematic duplicate logging code smells, developers considered them as technical debt and were reluctant to fix. For the rest of the potentially problematic instances, developers agreed that they are problematic and fixed them. For the justifiable ones, we communicated with developers for discussion (e.g., emails or posts on developers' forums).

We implemented a static analysis tool, \tool, to automatically detect {\em problematic} duplicate logging code smells. \toolS leverages the findings from our manual study, including the uncovered patterns of duplicate logging code smells and the categorization on problematic and justifiable cases.
We evaluated \toolS on eight systems: five are from the manual study and three are additional systems. We also applied \toolS on the updated versions of the five manually studied systems. The evaluation shows that the uncovered patterns of the duplicate logging code smells also exist in the additional systems, and duplicate logging code smells may be introduced over time. An automated approach such as \toolS can help developers avoid duplicate logging code smells as systems evolve.
In total, we reported 91 instances of duplicate logging code smell to developers and all the reported instances are fixed. %Our fix suggestions are all accepted by developers.
%\footnote{\label{link}The link to share our archived data is omitted due to the double blind review policy.}
Figure~\ref{fig:overall} shows the overall process of finding and detecting duplicate logging code smells.

%Moreover, during the process of our manual study, we found that duplicate logging statements might be a consequence \zhenhao{or side effect?} of code clones.
%To provide a more comprehensive understanding of duplicate logging code smells, we further investigate the relationship between the problematic instances of duplicate logging code smells and code clones. 
We further investigate the relationship between the problematic instances of duplicate logging code smells and code clones. Intuitively, duplicate logging statements could be related to, or are even a consequence of code clones (e.g., logging statements can be copied along with other code since cloning is often performed hastily without much attention on the context~\cite{5463343}). 
The findings of our study may show other negative effect of code clones on logging statements and inspire future code clone and logging research. 
%The findings of our study may show other negative effect of code clones on non-functional code (i.e., logging statements) and may inspire future code clone and logging research. 
We combine both an automated code clone detection tool (i.e., NiCad~\cite{nicad}) and manual study on the eight studied systems to examine if the duplicate logging code smell instances reside in cloned code snippets. We find that 83\% of the problematic duplicate logging code smell instances reside in cloned code snippets; however, 17\% of the instances reside in very short code blocks that are difficult to detect using automated code detection tools. %We match the locations of the code clones with the problematic instances of duplicate logging code smells. %To mitigate possible false positives in the clone detection tool, we then manually study the 

In summary, this paper makes the following contributions: %that we receive into a static code checker that we implement. %Once the anti-patterns are derived, we then manually study the genealogy
%~\jinqiu{The last bullet can be the third one? The findings first, then the tool and result.}
\vspace{-0.1cm}
\begin{itemize} \itemsep 0em
  \item We uncovered five patterns of duplicate logging code smells through an extensive manual study on over 4K duplicate logging statements.%This is the first study on helping developers refactor logging code.

  %\item We find that although 60\% of duplicate logs are related to duplicate code, 40\% of them are not.
  \item We presented a categorization of duplicate logging code smells (i.e., problematic or justifiable), based on both our manual analysis %(i.e., studying the logging statement and its surrounding code) 
  and developers' feedback.

%  \item We manually assessed every logging code smell instance, whether it is problematic or justifiable, based our understanding (i.e., studying the logging statement and its surrounding code) and developers' feedback. %We further categorize each pattern of duplicate logging code smell into sub-categories based on its potential impact.

  \item We proposed \tool, a static analysis tool that integrates our manual study result and developers' feedback to detect problematic duplicate logging code smells. %We evaluated \toolS for both the accuracy and generalization (i.e., on new systems and on the newer versions as systems evolve).

  \item We reported 91 instances of problematic duplicate logging code smells to developers (\toolS is able to detect 81 of them), and all of them are fixed.

  %\item \peter{revisit this point}We found that almost half of the duplicate logging statements reside in cloned code snippets, and many of such clones reside in very short code blocks. %current clone detection tools might not be effective in detecting duplicate logging code smells.

  \item We found that most of the problematic instances of duplicate logging code smells (83.0\%) reside in cloned code snippets, which indicates that code clones may also result in bad logging practices that increase maintenance difficulties.

  \item We found that more than half of the duplicate logging statements reside in cloned code snippets, and a large portion of them reside in short code blocks (i.e., micro-clones) which are difficult to detect using existing code clone detection tools.

  %\item We find that considering duplicate logging statements help clone detection tools improve detection accuracy by 25\% to 47\%. 

  \item We found that duplicate logging statements have a non-negligible impact on helping the detection of code clones. After removing them, from 25.0\% to 47.1\% of the cloned code snippets with duplicate logging statements can not be detected as cloned code snippets. 

  \item We provided a replication package of our paper for future studies on logging code and code clones\footnote{We share the data of this paper at: \url{https://github.com/SPEAR-SE/Duplicate_Logs_Data}}.

  %\item Based on our manual analysis,\zhenhao{Keep this finding or disgard?} we provided a detailed discussion regarding the possibility of considering duplicate logging statements as an indicator of short cloned code snippets, and the potential of using duplicate logging statements to further improve clone detection tools.

\end{itemize}

Our study provides an initial step on creating a logging guideline for developers to improve the quality of logging code. \toolS is also able to detect duplicate logging code smells with high precision and recall. Future code clone studies should also consider other possible side effects of code clones (e.g., understanding system runtime behaviour), and may consider including information from other software artifacts (e.g., duplicate logging statements) to further improve clone detection results.

This work extends our previous work~\cite{DLFinder}. First, we add one more system to our manual study and extend our evaluation to include an additional system and compare our text-analysis-based algorithm on detecting inconsistently updated log messages with two baselines. We also add discussions on duplicate logging statements that do not belong to one of the uncovered smells.
Second, we study the relationship between duplicate logging statements, including the problematic instances of duplicate logging code smells, and code clones 
Finally, we investigate the potential impact between duplicate logging statements and code clones.

%\jinqiu{Based on RQ4 and RQ5, we can conclude that clone detection tools are not effective in detecting problematic duplicate logging statements. Might worth mentioning somehwere to highlight the necessity of our tool (however the ICSE paper is published already, seems no need to justify this??)}
\phead{Paper organization.} 
Section~\ref{sec:prestudy} describes data preparation %how we prepare the data for manual study (i.e., duplicate logging statements) 
and the studied systems. Section~\ref{sec:manual} discusses the process and the results of our manual study. %, and also developers' feedback on our results. 
Section~\ref{sec:detection} discusses the implementation details of \tool. Section~\ref{sec:results} presents the case study results. 
Section~\ref{sec:clone} investigates the relationship between problematic instances of duplicate logging code smells and code clones. Section~\ref{sec:rq5} investigates the relationship between duplicate logging statements and code clones, as well as the potential impact of duplicate logging statements on detecting code clones. Section~\ref{sec:threats} discusses the threats to validity of our study. Section~\ref{sec:related} surveys related work. Section~\ref{sec:conclusion} concludes the paper. %Appendix~\ref{sec:appendix1} analyzes how many duplicate logging statements (i.e., regardless whether or not they belong to one of the uncovered duplicate logging code smells) reside in cloned code snippets. 
Appendix~\ref{sec:appendix2} discusses the false positive rate of the automated clone detection tool. 
%Appendix~\ref{sec:appendix3} investigates the potential impact of duplicate logging statements on detecting code clones.

%% file: texfiles/prestudy.tex
%\section{Preliminary Study}
\section{Identifying Duplicate Logging Statements for Manual Study}
\label{sec:prestudy}
%In this section, we describe how we define duplicate logging statements and how we identify them for conducting a manual study. We also introduce the studied systems. % from which duplicate logging statements will be studied.

\phead{Definition and how to identify duplicate logging statements.} %\ian{phead inconsistent, only first letter capital or every first letter of each word capital?}
We define duplicate logging statements as logging statements that have identical static text messages. We focus on studying the log message because such semantic information is crucial for log understanding and system maintenance~\cite{Shang:2014:ULL:2705615.2706065, Yuan:2012:CLP:2337223.2337236}. As an example, the two following logging statements are considered duplicate: ``{\em Unable to create a new ApplicationId in SubCluster}" {\em + \textbf{ subClusterId.getId()}}, and ``{\em Unable to create a new ApplicationId in SubCluster}" {\em + \textbf{ id}}.

To prepare for a manual study, we identify duplicate logging statements by analyzing the source code using static analysis. In particular, the static text message of each logging statement is built by concatenating all the strings (i.e., constants and values of string variables) and abstractions of the non-string variables.
We also extract information to support the manual analysis, such as the types of variables that are logged, and the log level (i.e., {\em fatal}, {\em error}, {\em warn}, {\em info}, {\em debug}, or {\em trace}). Log levels can be used to reduce logging overheads in production (e.g., only record {\em info} and more severe levels) and may target different phases of software maintenance (e.g., {\em debug} logs may be used for debugging and {\em info} logs may provide information for general audience)~\cite{Li2017, Yuan:2012:CLP:2337223.2337236}. If two or more logging statements have the same static text message, they are identified as duplicate logging statements. 
%We exclude logging statements with only one word since those logging statements usually do not contain much static information, and are usually used to record the value of a dynamic variable during system execution.

\begin{table}
    \caption{An overview of the studied systems. }
    \vspace{-0.2cm}
    \centering
    %\resizebox{\textwidth}{!} {
    \tabcolsep=6pt
    \begin{tabular}{lrrrrrrr}
        \toprule
        \textbf{System}& \textbf{Version}  & \textbf{LOC} & \textbf{NOL} &
        \textbf{NODL} & \textbf{NODS} \\
        %&&&&&&& \textbf{logs} & \textbf{logs}\\
        \midrule
        \textbf{Cassandra} & 3.11.1  & 358K & 1.6K &  113 (7\%)& 46 \\
        \textbf{CloudStack} & 4.9.3  & 1.18M & 11.7K &  2.3K (20\%)& 865 \\
        \textbf{Elasticsearch} & 6.0.0  & 2.12M & 1.7K & 94 (6\%)& 40 \\
        \textbf{Flink} & 1.7.1  & 177K & 2.5K &  467 (11\%)& 203 \\
        \textbf{Hadoop} & 3.0.0  & 2.69M & 5.3K & 496 (9\%)& 217 \\

        \midrule
        \textbf{Camel} & 2.21.1  & 1.68M & 7.3K &  2.3K (32\%)& 886 \\
        \textbf{Kafka} & 2.1.0  & 542K & 1.5K &  406 (27\%)& 104 \\
        \textbf{Wicket} & 8.0.0 & 381K & 0.4K &  45 (11\%)& 21 \\
        % cass non-dup median 7, 0.146 neg
        % cloud stack non-dup median 8, 0.045 neg
        % es non-dup median 7, 0.27 small
        % hadoop non-dup median 6, 0.08 neg
        \bottomrule
    \end{tabular}

    \noindent LOC: lines of code, NOL: number of logging statements, NODL: number of duplicate logging statements, NODS: number of duplicate logging statements sets.
    \vspace{-0.6cm}
    %}
    \label{table:systems}
\end{table}
\phead{Studied systems.}
Table~\ref{table:systems} shows the statistics of the studied systems. We identify duplicate logging statements from the top five large-scale open source Java systems in the table for our manual analysis: Hadoop, CloudStack, Elasticsearch, Cassandra and Flink which are commonly used in prior studies for log-related research~\cite{Li2018, log_pattern_ICSE2017, DLFinderSRC,mehran_emse_2018,nemo_log_issue}. The studied systems also use popular Java logging libraries~\cite{nemo_logging_utility} (e.g., Log4j~\cite{log4j} and SLF4J~\cite{SLF4J}). Hadoop is a distributed computing framework, CloudStack is a cloud computing platform, Elasticsearch is a distributed search engine, Cassandra is a NoSQL database system, and Flink is a stream-processing framework. These systems belong to different domains and are well maintained. We study all Java source code files in the main branch of each system and exclude test files, since we are more interested in studying duplicate logging statements that may affect log understanding in production. In general, we find that there is a non-negligible number of duplicate logging statements in the studied systems (6\% to 20\%). 
%The median number of words in the duplicate logging statements are similar to that of non-duplicate logging statements (i.e., both range from 6 to 8 words). 
%The median number of words in the duplicate logging statements are similar to that of non-duplicate logging statements (i.e., both range from 6 to 8 words), which shows that they have a similar level of semantic information (in terms of the number of words).

%% file: texfiles/pattern.tex
\section{Patterns of Duplicate Logging Code Smells}
\label{sec:manual}

In this section, we conduct a manual study to investigate duplicate logging statements. Note that duplicate logging statements may not necessarily be a problem. Hence, our goal is to uncover patterns of potential code smells that may be associated with duplicate logging statements (i.e., {\em duplicate logging code smells}). % and discuss other duplicate logging statements that may not be a problem. 
Similar to prior code smell studies, we consider duplicate logging code smells as a {\em ``surface indication that usually corresponds to a deeper problem in the system''}~\cite{budgen2003software, fowler1999refactoring}. Such duplicate logging code smells may be indications of logging problems that require fixes.
%In this section, we conduct a manual study to uncover patterns of potential code smells that may be associated with duplicate logging statements (i.e., {\em duplicate logging code smells}). Similar to prior code smell studies, we consider duplicate logging code smells as a {\em ``surface indication that usually corresponds to a deeper problem in the system''}~\cite{budgen2003software, fowler1999refactoring}. Such duplicate logging code smells may be indications of logging problems that require fixes.

We categorize each duplicate logging code smell instance as either problematic (i.e., require fixes) or justifiable (i.e., do not require fixes), by understanding the surrounding code.
Not every duplicate logging code smell is problematic. Intuitively, one needs to consider the code context to decide whether a code smell instance is problematic and requires fixes. As shown in prior studies~\cite{Zhu:2015:LLH:2818754.2818807, Fu:2014:DLE:2591062.2591175, Li2018}, logging decisions, such as log messages and log levels, are often associated with the structure and semantics of the surrounding code.
In addition to the manual analysis by the authors, we also ask for developers' feedback regarding both the problematic and justifiable cases.
%Hence, we categorize code smell instances as problematic or justifiable through both our manual analysis on surrounding code and also communications with developers.
By providing a more detailed understanding of code smells, we may better assist developers to improve logging practices and inspire future research.

\phead{Manual study process.} We conduct a manual study by analyzing all the duplicate logging statements in the five studied systems. In total, we studied 1,371 sets of duplicate logging statements (more than 4K logging statements in total; each set contains two or more logging statements with the same static message). Specifically, we examine the four following criteria when studying the code snippets: 1) the generated log messages record incorrect information (i.e., the recorded method name is different from the method where the log message is generated), 2) the recorded information cannot be used to distinguish the occurred errors (e.g., to distinguish different exception types), 3) there are inconsistencies in terms of log levels or the recorded debugging information, and 4) the duplicated log message may need to be updated to ensure consistency (i.e., maintenance of logs).

\noindent The process of our manual study involves five phases:

%\vspace{-0.1cm}
%\begin{itemize}\itemsep 0em
{\em Phase I}: The first two authors manually studied 301 randomly sampled (based on 95\% confidence level and 5\% confidence interval~\cite{boslaugh2008statistics}) sets of duplicate logging statements and the surrounding code to derive an initial list of duplicate logging code smell patterns. All disagreements were discussed until a consensus was reached.

%The authors studied all of the static messages, log levels, and surrounding code/comments to derive the patterns.
{\em Phase II}: The first two authors {\em independently} categorized {\em all} of the 1,371 sets of duplicate logging statements to the derived patterns in Phase I. We did not find any new patterns in this phase. The results of this phase have a Cohen’s kappa of 0.811, which is a substantial-level of agreement~\cite{kappa}.

{\em Phase III}: The first two authors discussed the categorization results obtained in Phase II. All disagreements were discussed until a consensus was reached.

{\em Phase IV}: The first two authors further studied all logging code smell instances that belong to each pattern to identify justifiable cases that may not need fixes. The instances that do not belong to the category of justifiable are considered potentially problematic and may require fixes.

{\em Phase V}: We verified both the problematic and justifiable instances of logging code smells with developers by creating pull requests, sending emails, or posting our findings on developers’ forums (e.g., Stack Overflow). We reported every instance that we believe to be problematic (i.e., require fixes), and reported a number of instances for each justifiable category.

\begin{table}
\caption{Patterns of duplicate logging code smells
and corresponding examples.}
\vspace{-0.2cm}
\centering
\resizebox{\columnwidth}{!} {%
%\begin{adjustbox}{width=1\textwidth}
\scalebox{1}{
\begin{tabular}{m{0.04\textwidth} | m{.60\textwidth}  }%{ m{3cm} |l}
\toprule
\textbf{Pattern} & \textbf{Example} \\
%%%%%%%%%%%%%%%%%%%%%%%%%%%%%%%%%
\midrule
%This example is from Cloudstack, com.cloud.api.dispatch.ParamProcessWorker.processParameters
IC &
\includegraphics[width=0.60\textwidth]{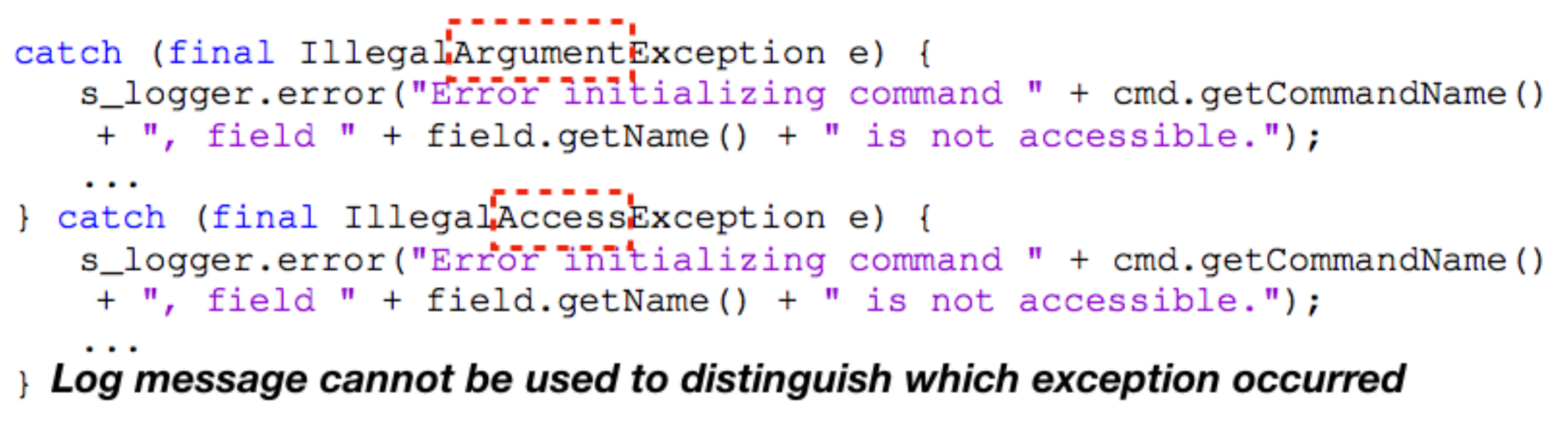}

\\
%%%%%%%%%%%%%%%%%%%%%%%%%%%%%%%%%
\midrule
%This example is from Cloudstack
IE & \includegraphics[width=0.60\textwidth]{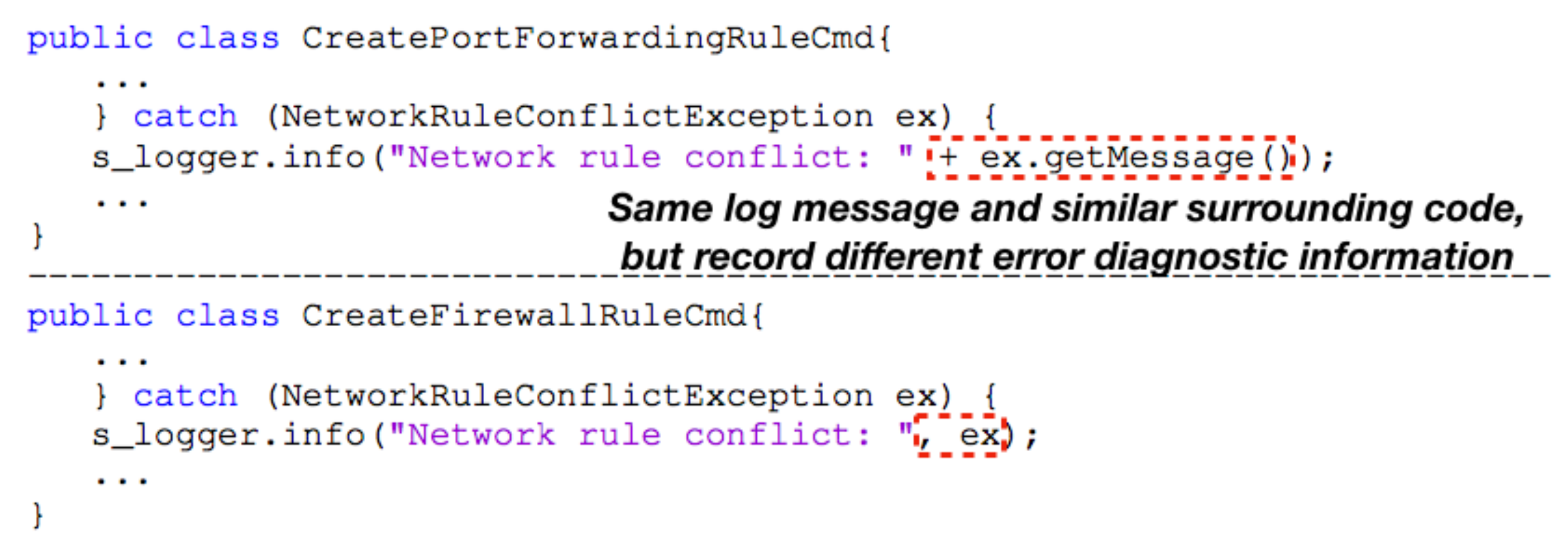}
\\
%%%%%%%%%%%%%%%%%%%%%%%%%%%%%%%%%
\midrule
%This is from CloudStack
LM & \includegraphics[width=0.60\textwidth]{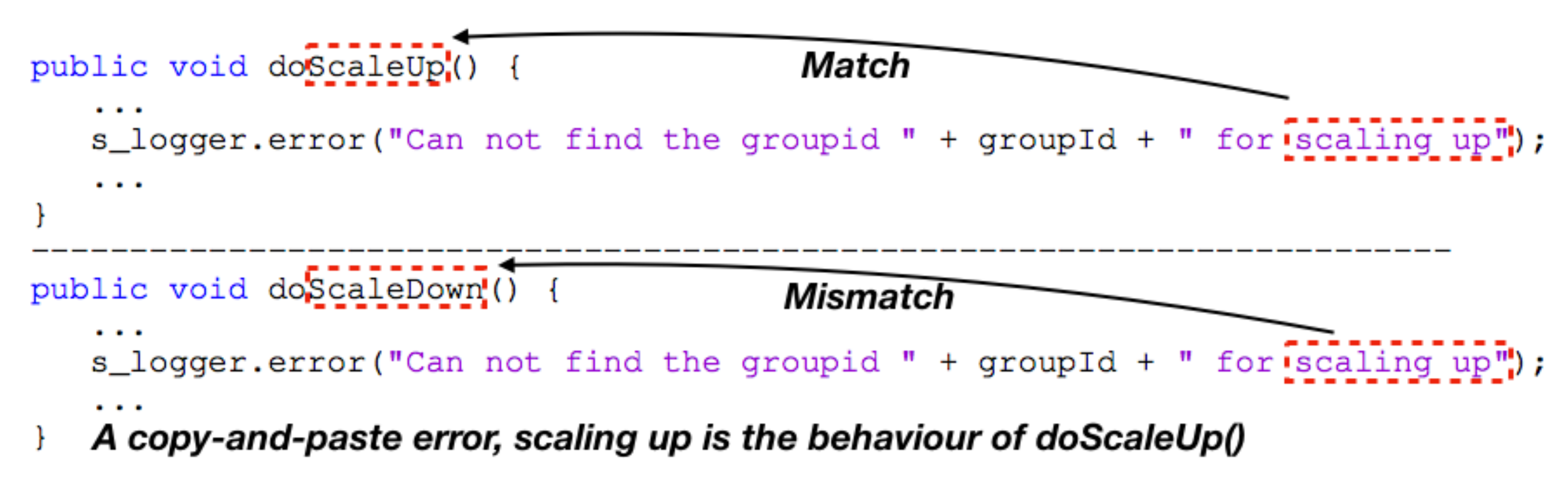}
\\
%%%%%%%%%%%%%%%%%%%%%%%%%%%%%%%%%

\midrule
%This is from Cassandra, same class: CompactionManager
IL & \includegraphics[width=0.60\textwidth]{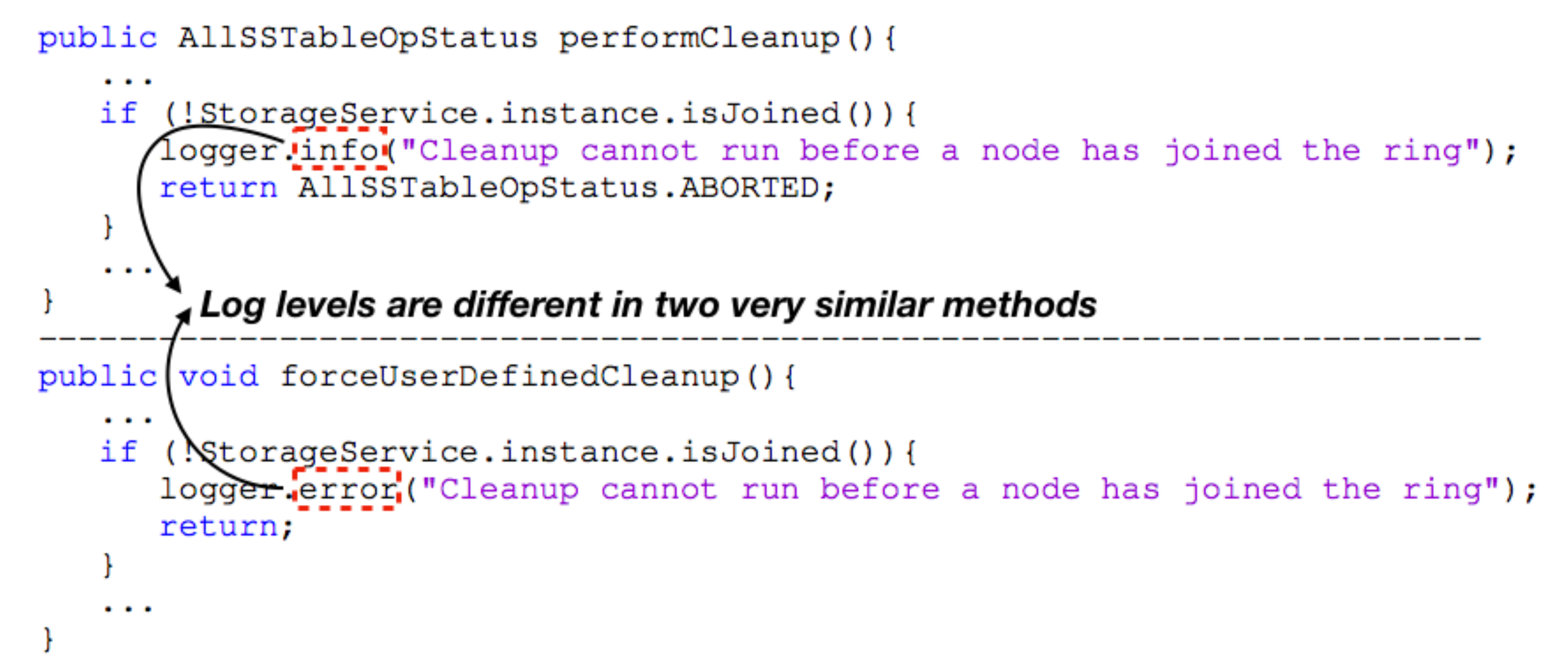}
\\

%%%%%%%%%%%%%%%%%%%%%%%%%%%%%%%%%

\midrule
%This example is from Hadoop
DP & \includegraphics[width=0.60\textwidth]{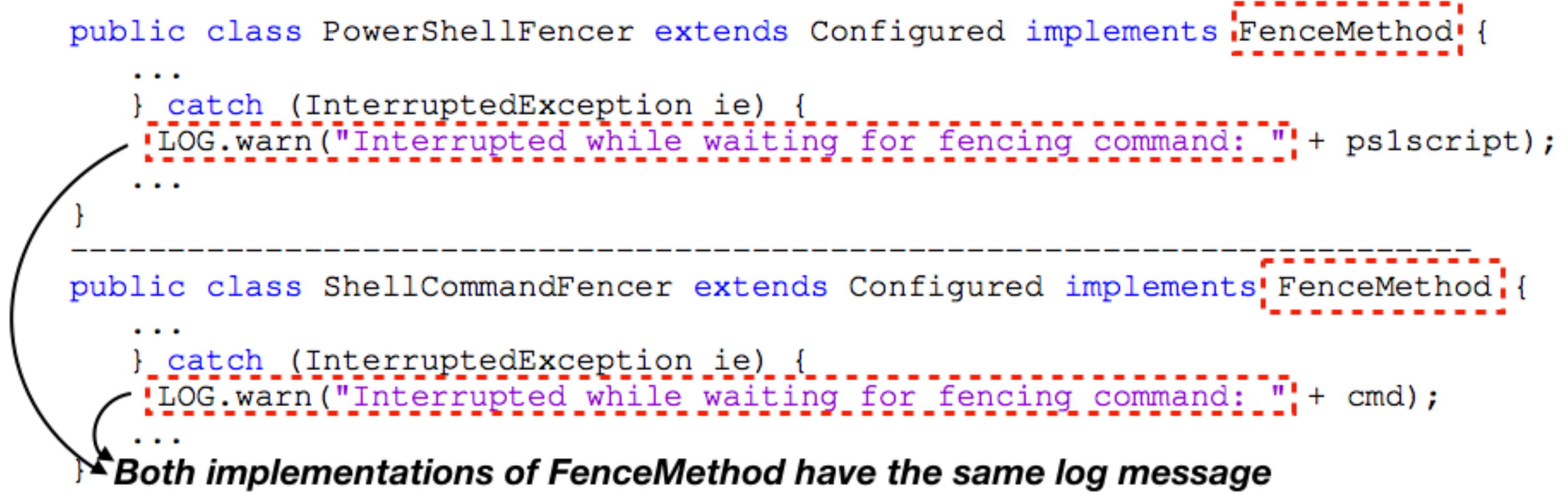}
\\

%%%%%%%%%%%%%%%%%%%%%%%%%%%%%%%%%

\bottomrule
\end{tabular}}
%\vspace{-0.3cm}
}
%\end{adjustbox}
\vspace{-0.3cm}
\label{tab:patterns}
\end{table}

\begin{table}
    \caption{Number of problematic instances ({\sf Prob.}) verified by our manual study and developers' feedback, number of instances of technical debt ({\sf Tech.}), and total number of instances (Total) including non-problematic instances.\label{tab:manual}}
	\vspace{-0.3cm}
    \centering
    %\resizebox{\columnwidth}{!} {
    \tabcolsep=1pt
    \begin{tabular}{lcc|cc|cc|cc|cc}

        \toprule
        & \multicolumn{2}{c|}{\textbf{IC}} & \multicolumn{2}{c|}{\textbf{IE}} & \multicolumn{2}{c|}{\textbf{LM}} & \multicolumn{2}{c|}{\textbf{IL }} & \multicolumn{2}{c}{\textbf{DP}}\\
        & Prob.  & Total & Prob. &Total& Prob.  &Total& Prob. &Total& Tech.  &Total\\
        \midrule
        \textbf{Cassandra}     & 1  & 1 & 0  & 1 & 0  & 0 & 0 & 3 & 2  & 2 \\
        \textbf{CloudStack}    & 8  & 8 & 4  & 14 & 27  & 27 & 0 & 47 & 107 & 107 \\
        \textbf{Elasticsearch} & 1  & 1 & 0  & 5 & 1  & 1 & 0 & 9 & 3  & 3 \\
        \textbf{Flink}        & 0  & 0 & 2  & 5 & 4  & 4 & 0 & 14 & 24  & 24 \\
        \textbf{Hadoop}        & 5  & 5 & 0  & 0 & 9  & 9 & 0 & 17 & 27  & 27 \\

        \midrule
        \textbf{Total}         & 15  & 15 & 6 & 25 & 41  & 41 & 0  & 90 & 163\tnote{1}  & 163 \\

        \bottomrule
    \end{tabular}

	\vspace{-0.6cm}

    %}
\end{table}

\phead{Results.} In total, we uncovered five patterns of duplicate logging code smells. Table~\ref{tab:patterns} lists the uncovered code smell patterns and the corresponding examples. Table~\ref{tab:manual} shows the number of problematic code smell instances for each pattern that we manually found.
Below, we discuss each pattern according to the following template:

\begin{LaTeXdescription}
  \item[{\em Description:}] A description of the pattern of duplicate logging code smell.
  \item[{\em Example:}] An example of the pattern.
  \item[{\em Code smell instances:}] Discussions on the manually-uncovered code smell instances. We also discuss the justifiable cases if we found any.
  \item[{\em Developers' feedback:}] A summary of developers' feedback on both the problematic and justifiable cases.

\end{LaTeXdescription}

\phead{Pattern 1: Inadequate information in catch blocks (IC).}

\noindent{\bf {\em Description. }}Developers usually rely on logs for error diagnostics when exceptions occur~\cite{Yuan:2014:STP:2685048.2685068}. However, we find that sometimes, duplicate logging statements in different {\em catch} blocks of the same {\em try} block may cause debugging difficulties since the logs fail to tell which exception occurred.

\noindent{\bf {\em Example. }}As shown in Table~\ref{tab:patterns}, in the {\tt\small ParamProcessWorker} class in CloudStack, the {\em try} block contains two {\em catch} blocks; however, the log messages in these two {\em catch} blocks are identical. Since both the exception message and stack trace are not logged, once one of the two exceptions occurs, developers may encounter difficulties in finding the root causes and determining the occurred exception.

\noindent{\bf {\em Code smell instances. }}After examining all the instances of IC, we find that all of them are potentially problematic and require fixes. For all the instances of IC, none of the exception type, exception message, and stack trace are logged. %In other words, developers may not be able to identify what the occurred exception is when analyzing the logs, since the log messages are the same and there is insufficient error diagnostic information (e.g., no stack trace nor exception message) being recorded by the logging statements.

\noindent{\bf {\em Developers' feedback. }}We reported all the problematic instances of IC (15 instances), and all of them are fixed by adding more error diagnostic information (e.g., stack trace) into the logging statements. %  All the pull requests were accepted by developers and the fixes were integrated to the studied systems. 
Developers agree that IC will cause confusion and insufficient information in the logs, which may increase the difficulties of error diagnostics.

\phead{Pattern 2: Inconsistent error-diagnostic information (IE).}
\noindent{\bf {\em Description. }}We find that sometimes duplicate logging statements for recording exceptions may contain inconsistent error-diagnostic information (e.g., one logging statement records the stack trace and the other does not), even though the surrounding code is similar. %Namely, the recorded dynamic variables are different (e.g., one log records the stack trace but the other log does not), .

\noindent{\bf {\em Example. }}As shown in Table~\ref{tab:patterns}, the two classes in CloudStack: {\tt\small Create\textbf{PortForwarding}RuleCmd} and {\tt\small Create\textbf{Firewall}RuleCmd} have similar functionalities. The two logging statements have the same static text message and are in methods with identical names (i.e., {\em create()}, not shown due to space restriction). %The two {\em create()} methods are very similar to each other in terms of code structure.
The {\em create()} method in {\tt\small Create\textbf{PortForwarding}RuleCmd} is about creating rules for port forwarding, and the method in {\tt\small Create\textbf{Firewall}RuleCmd} is about creating rules for firewalls. These two methods have very similar code structure and business logic. However, the two logging statements record different information: One records the stack trace information and the other one only records the exception message (i.e., {\em ex.getMessage()}). Since the two logging statements have similar context, the error-diagnostic information recorded by the logs may need to be consistent for the ease of debugging. We reported this example, which is now fixed to have consistent error-diagnostic information.

\noindent{\bf {\em Code smell instances. }} We find 25 instances of IE (Table~\ref{tab:manual}), and six of them are considered problematic in our manual study. %(all pull requests are accepted by developers).
From the remaining instances of IE, we find three justifiable cases that may not require fixes.

{\em \underline{Justifiable case IE.1:} Duplicate logging statements record general and specific exceptions}.
For 11/25 instances of IE, we find that the duplicate logging statements are in the {\em catch} blocks of different types of exception. In particular, one duplicate logging statement is in the {\em catch} block of a generic exception (i.e., the {\tt\small Exception} class in Java) and the other one is in the {\em catch} block of a more specific exception (e.g., application-specific exceptions such as {\tt\small CloudRuntimeException}). In all of the 11 cases, we find that one log would record the stack trace for \texttt{\small Exception}, and the duplicate log would only record the type of the occurred exception (e.g., by calling {\em e.getMessage()}) for a more specific exception. The rationale may be that generic exceptions, once occurred, are often not expected by developers~\cite{Yuan:2014:STP:2685048.2685068}, so it is important that developers record more error-diagnostic information.

{\em \underline{Justifiable case IE.2:} Duplicate logging statements are in the same catch block for debugging purposes}.
For 6/25 instances of IE, the duplicate logging statements are {\em in the same} {\em catch} block and developers' intention is to use a duplicate logging statement at {\em debug} level to record rich error-diagnostic information such as stack trace (and the log level of the other logging statement could be {\em error}). The extra logging statements at {\em debug} level help developers debug the occurred exception and reduce logging overhead in production~\cite{Li2017} (i.e., logging statements at debug level are turned off). %In particular, the duplicate log at the {\em debug} level records additional information (e.g., stack trace) that helps developers debug the occurred exception and reduces logging overhead in production~\cite{Li2017}.

{\em \underline{Justifiable case IE.3:} Having separate error-handling classes}. For 2/25 instances, we find that the error-diagnostic information is handled by creating an object of an error-handling class. As an example from CloudStack:

\begin{lstlisting}
public final class LibvirtCreateCommandWrapper {
    ...
        } catch (final CloudRuntimeException e) {
            s_logger.debug("Failed to create volume: " + e.toString());
            return new CreateAnswerErrorHandler(command, e);
        }
    ...
}

public class KVMStorageProcessor {
    ...
        } catch (final CloudRuntimeException e) {
            s_logger.debug("Failed to create volume: ", e);
            return new CopyCmdAnswerErrorHandler(e.toString());
        }
    ...

}
\end{lstlisting}
In this example, extra logging is added by using error-handling classes (i.e., {\tt\small CreateAnswerErrorHandler} and {\tt\small CopyCmdAnswerErrorHandler}) to complement the logging statements. As a consequence, the {\em actual} logged information is consistent in these two methods: One method records {\em e.toString()} in the logging statement and records the exception variable {\em e} through an error-handling class; the other method records {\em e} in the logging statement and records {\em e.toString()} through an error-handling class.

\noindent{\bf {\em Developers' feedback. }} We reported all the six instances of IE that we consider problematic to developers, all of which are fixed. Moreover, we ask developers whether our conjecture was correct for each of the justifiable cases of IE. Developers confirmed our observation on the justifiable cases. They agreed that those cases are not problematic thus do not require fixes. 
%We received positive feedback that confirms our manual analysis on the justifiable cases.

\phead{Pattern 3: Log message mismatch (LM).}

\noindent{\bf {\em Description. }}Sometimes after developers copy and paste a piece of code to another method or class, they may forget to change the log message. This results in having duplicate logging statements that record inaccurate system behaviors.

%resulting in duplicate logging statements in different code locations.

\noindent{\bf {\em Example. }} As an example, in Table~\ref{tab:patterns}, the method {\em doScaleDown()} is a code clone of {\em doScaleUp()} with very similar code structure and minor syntactical differences. However, developers forgot to change the log message in {\em doScaleDown()}, after the code was copied from {\em doScaleUp()} (i.e., both log messages contain {\em scaling up}). Such instances of LM may cause confusion when developers analyze the logs.

\noindent{\bf {\em Code smell instances. }} We find that there are 41 instances of LM that are caused by copying-and-pasting the logging statement to new locations without proper modifications. For all the 41 instances, the log message contains words of incorrect class or method name that may cause confusion when analyzing logs.

\noindent{\bf {\em Developers' feedback. }} Developers agree that the log messages in LM should be changed in order to correctly record the execution behavior (i.e., update the copy-and-pasted log message to contain the correct class/method name). We reported all the 41 instances of LM that we found and all of them are fixed. %through pull requests, and all of the reported instances are now fixed.

\phead{Pattern 4: Inconsistent log level (IL).}
%76 instances in total among the 4 systems.

\noindent{\bf {\em Description. }}Log levels (e.g., {\em fatal}, {\em error}, {\em info}, {\em debug}, or {\em trace}) allow developers to specify the verbosity of the log message and to reduce logging overhead when needed% (e.g., {\em debug} is usually disabled in production)
~\cite{Li2017}. A prior study~\cite{Yuan:2012:CLP:2337223.2337236} shows developers frequently modify log levels to find the most adequate level. We find that there are duplicate logging statements that, even though the log messages are exactly the same, the log levels are different.

\noindent{\bf {\em Example. }} In the IL example shown in Table~\ref{tab:patterns}, the two methods, which are from the same class {\tt\small CompactionManager}, have very similar functionality (i.e., both try to perform cleanup after compaction), but different log levels. %we find that the log levels are different in these two methods.

\noindent{\bf {\em Code smell instances. }} We find three justifiable cases in IL that may be developers' intended behavior. We do not find problematic instances of IL after communicating with developers -- Developers think the problematic instances identified by our manual analysis may not be problems.

{\em \underline{Justifiable case IL.1:} Duplicate logging statements are in the catch blocks of different types of exception}. Similar to what we observed in IE, we find that for 9/90 instances, the log level for a more generic exception is usually more severe (e.g., {\em error} level for the generic Java {\tt\small Exception} and {\em info} level for an application-specific exception). Generic exceptions might be unexpected to developers~\cite{Yuan:2014:STP:2685048.2685068}, so developers may use a higher log level (e.g., {\em error}) to record exception messages.

{\em \underline{Justifiable case IL.2:} Duplicate logging statements are in different branches of the same method}. There are 42/90 instances belong to this case. Below is an example from Elasticsearch, where a set of duplicate logging statements occur in the same method but in different branches.
\begin{lstlisting}
   if (lifecycle.stoppedOrClosed()) {
        logger.trace("failed to send ping transport message", e);
   } else {
        logger.warn("failed to send ping transport message", e);
   }
\end{lstlisting}
In this case, developers already know the desired log level and intend to use different log levels due to the difference in execution (i.e., in the if-else block). %Note that the condition may also be a log level guard (e.g., {\tt if (LOGGER.isDebugEnabled()}) or a {\tt switch} statement.

{\em \underline{Justifiable case IL.3:} Duplicate logging statements are followed by error-handling code}. There are 19/90 instances that are observed to have such characteristics: In a set of duplicate logging statements, some statements have log levels of higher verbosity, and others have log levels of lower verbosity. However, the duplicate logging statement with lower verbosity log level is followed by additional error handling code (e.g., {\em throw a new Exception(e);}). Therefore, the error is handled elsewhere (i.e., the exception is re-thrown), and may be recorded at a higher-verbosity log level.

\noindent{\bf {\em Developers' feedback. }} In all the instances of IL that we found, developers think that IL may not be a problem. In particular, developers agreed with our analysis on the justifiable cases. However, developers think the problematic instances of IL from our manual analysis may also not be problems. We concluded the following two types of feedback from developers on the ``suspect'' instances of IL (i.e., 20 problematic ones from our manual analysis out of the 90 instances of IL).
The first type of developers' feedback argues the importance of semantics and usage scenario of logging in deciding the log level. A prior study~\cite{Yuan:2012:CLP:2337223.2337236} suggests that logging statements that appear in syntactically similar code, but with inconsistent log levels, are likely problematic. However, based on the developers' feedback that we received, IL still may not be a concern, even if the duplicate logging statements reside in very similar code. A developer indicated that ``conditions and messages are important but the {\it context} is even more important''. As an example, both of the two methods may display messages to users. One method may be displaying the message to {\em local} users with a {\em debug} logging statement to record failure messages. The other method may be displaying the message to {\em remote} users with an {\em error} logging statement to record failure messages (problems related to remote procedure calls may be {\em more severe} in distributed systems). Hence, even if the code is syntactically similar, the log level has a reason to be different due to the different semantics and purposes of the code (i.e., referred to as different {\it contexts} in developers' responses). Our findings show that future studies should consider both the syntactic structure and semantics of the code when suggesting log levels.

The second type of developers' feedback acknowledges the inconsistency. However, developers are reluctant to fix such inconsistencies since developers do not view them as concerns.
For example, we reported the instance of IL in Table~\ref{tab:patterns} to developers. A developer replied:
\noindent ``I think it should probably be an {\it ERROR} level, and I missed it in the review (could make an argument either way, I do not feel strongly that it should be {\it ERROR} level vs {\it INFO} level.'' %, but it should be consistent).''
Our opinions (i.e., from us and prior studies~\cite{Yuan:2012:CLP:2337223.2337236, Li2017}) differ from that of developers' regarding whether such inconsistencies are problematic.
On one hand, whether an instance of IL is problematic or not can be subjective. This shows the importance of including perspectives from multiple parties (e.g., user studies or interviews) in future studies of software logging practice.
On the other hand, the discrepancy also indicates the need of establishing a guidance for logging practice and further even enforcing such standard. In short, none of the IL instances that we manually identified are problematic based on developers' feedback.

\phead{Pattern 5: Duplicate logging statements in polymorphism (DP).}

\noindent{\bf {\em Description. }}Classes in object-oriented languages are expected to share similar functionality if they inherit the same parent class or if they implement the same interface (i.e., polymorphism). Since log messages record a higher level abstraction of the program~\cite{Shang:2014:ULL:2705615.2706065}, we find that even though there are no clones among a parent method and its overridden methods, such methods may still contain duplicate logging statements. Such duplicate logging statements may cause maintenance overhead. For example, when developers update one log message, they may forget to update the log message in all the other sibling classes. Inconsistent log messages may cause problems during log analysis~\cite{mehran_emse_2018, HADOOP-4190, logzip, loghub}.

\noindent{\bf {\em Example. }} In Table~\ref{tab:patterns}, the two classes ({\tt\small PowerShellFencer} and {\tt\small ShellCommandFencer}) in Hadoop both extend the same parent class, implement the same interface, and share similar behaviors. The inherited methods in the two classes have identical log message. However, as the system evolves, developers may not always remember to keep the log messages consistent, which may cause problems during system debugging, understanding, and analysis.

\noindent{\bf {\em Code smell instances. }} We find that all the 163 instances of DP are potentially problematic that may be fixed by refactoring. In most of the instances, the parent class is an abstract class, and the duplicate logging statements exist in the overridden methods of the subclasses. We also find that in most cases, the overridden methods in the subclasses are very similar with minor differences (e.g., to provide some specialized functionality), which may be the reason that developers use duplicate logging statements.

\noindent{\bf {\em Developers' feedback. }} Developers agree that DP is associated with logging code smells and specific refactoring techniques are needed. %From the feedback that we received, developers generally agree that DP is related to logging code smells.
One developer comments that:

\noindent {\em ``You want to care about the logging part of your code base in the same way as you do for business-logic code (one can argue it is part of it), so salute DRY (do-not-repeat-yourself).''}

Based on developers' feedback, DP is viewed more as technical debts~\cite{Kruchten:2012:TDM:2412381.2412847}, while resolving DP often requires systematic refactoring. 
However, to the best of our knowledge, current Java logging frameworks, such as SLF4J and Log4j 2, do not support the use of polymorphism in logging statements. Thus, we find that developers are more reluctant to fix DP. The way to resolve DP is to ensure that the log message of the parent class can be reused by the subclasses, e.g., storing the log message in a static constant variable. We received similar suggestions from developers on how to refactor DP, such as {\it ``adding a method in the parent class that generates the error text for that case:  logger.error(notAccessible( field.getName()));}'', or {\it ``creat[ing] your own Exception classes and put message details in them''.} We find that without supports from logging frameworks, even though developers acknowledged the issue of DP, they do not want to {\em manually} fix the code smells. Similar to some code smells studied in prior research~\cite{Johnson:2013:WDS:2486788.2486877, Silva:2016:WWR:2950290.2950305}, developers may be reluctant to fix DP due to additional maintenance overheads but limited supports (i.e., need to manually fix hundreds of DP instances). Therefore, we did not report all the instances of DP and refer to the instances of DP as technical debts, instead of problematic instances, in the rest of the paper.  In short, logging frameworks should provide better support to developers in creating log ``templates'' that can be reused in different places in the code.  % that exceeds the benefits

\phead{Discussions on duplicate logging statements that do not belong to one of the uncovered smells.} In this paper, we focus on studying the problematic patterns of duplicate logging statements. However, we do not consider all duplicate logging statements as bad logging practice. For other duplicate logging statements that do not belong to the identified smells, we did not find evidence that they may cause confusion when analyzing logs. In most of the cases, the log message may be similar by coincidence (e.g., the log messages are used to record a certain type of exception message and stack trace). In some cases, we found that developers intentionally write duplicate logging statements with comments explaining the reasons. For example, some developers mentioned in the comment that the code snippet is copied from another class, and said the code should be refactored in the future. In some other cases, developers described the intention of the two duplicate logging statements. Although the static messages are identical, the comments are different, which shows that duplicate logging statements could have different intentions in different places. In such cases, duplicate logging statements may assist machine-learning based approaches to suggest where-to-log.

\rqboxc{We manually uncovered five patterns of duplicate logging code smells. % and six justifiable cases where the code smell instances may not need fixes. 
In total, our manual study helped developers fix 62 problematic duplicate logging code smells in the studied systems.}
\vspace{-0.2cm}

%% file: texfiles/detection.tex
\section{DLFinder: Automatically Detecting Problematic Duplicate Logging Code Smells}
\label{sec:detection}

Section~\ref{sec:manual} uncovers five patterns of duplicate logging code smells, and provides guidance in identifying {\em problematic} logging code smells. 
To help developers detect such problematic code smells and improve logging practices, we propose an automated approach, specifically a static analysis tool, called \tool. \toolS uses abstract syntax tree (AST) analysis, data flow analysis, and text analysis. Note that we exclude the detection result of IL (i.e., inconsistent log level) in this study, since based on the feedback from developers, none of the IL instances are problematic. Below, we discuss how \toolS detects each of the four patterns of duplicate logging code smell (i.e., IC, IE, LM, and DP).

\phead{Detecting inadequate information in catch blocks (IC).}
%\noindent{\bf {\em Detection Approach.}}
\toolS first locates the {\em try-catch} blocks that contain duplicate logging statements. Specifically, \toolS finds the {\em catch} blocks of the same {\em try} block that catch different types of exceptions, and these {\em catch} blocks contain the same set of duplicate logging statements. Then, \toolS uses data flow analysis to analyze whether the handled exceptions in the {\em catch} blocks are logged (e.g., record the exception message). \toolS detects an instance of IC if none of the logging statements in the {\em catch} blocks record either the stack trace or the exception message.

\phead{Detecting inconsistent error-diagnostic information (IE).}
%{\em Detection Approach.}
\toolS first identifies all the {\em catch} blocks that contain duplicate logging statements. Then, for each {\em catch} block, \toolS uses data flow analysis to determine how the exception is logged by analyzing the usage of the exception variable in the logging statement. Namely, the logging statement records 1) the entire stack trace, 2) only the exception message, or 3) nothing at all.
%If there is another log line with the same message in other \texttt{catch} blocks (i.e., duplicate log), \toolS would compare how the exception variable is logged.
Then, \toolS compares how the exception variable is used/recorded in each of the duplicate logging statements.
\toolS detects an instance of IE if a set of duplicate logging statements that appear in {\em catch} blocks has an inconsistent way of recording the exception variables (e.g., the log in one {\em catch} block records the entire stack trace, and the log in another {\em catch} block records only the exception message, while the two catch blocks handle the same type of exception). Note that for each instance of IE, the multiple {\em catch} blocks with duplicate logging statements in the same set may belong to different {\em try} blocks. In addition, \toolS decides if an instance of IE can be excluded if it belongs to one of the three justifiable cases (IE.1--IE.3) by checking the exception types, if the duplicate logging statements are in the same {\em catch} block, and if developers pass the exception variable to another method. %If so, the instance is marked as potentially justifiable, and thus, excluded by \tool. %Finally, \toolS identifies the inconsistent in calls to exception variables in duplicate logs.

\vspace{-0.1cm}
\phead{Detecting log message mismatch (LM)}. LM is about having an incorrect method or class name in the log message (e.g., due to copy-and-paste). Hence, \toolS analyzes the text in both the log message and the class-method name (i.e., concatenation of class name and method name) to detect LM by applying commonly used text analysis approaches~\cite{Chen:2016:SUT:2992358.2992444}. \toolS detects instances of LM using four steps: 1) For each logging statement, \toolS splits class-method name into a set of words (i.e., {\em name set}) and splits log message into a set of words (i.e., {\em log set}) by leveraging naming conventions (e.g., camel cases) and converting the words to lower cases. 2) \toolS applies stemming on all the words using Porter Stemmer~\cite{stemmer}. 3) \toolS removes stop words in the log message. We find that there is a considerable number of words that are generic across the log messages in a system (e.g., on, with, and process). Hence, we obtain the stop words by finding the top 50 most frequent words (our studied systems has an average of 3,352 unique words in the static text messages) across all log messages in each system~\cite{Yang:2014:SIS:2683115.2683138}. 4) For every logging statement, between the name set (i.e., from the class-method name) and its associated log set, \toolS counts the number of common words shared by both sets. Afterward, \toolS detects an instance of LM if the number of common words is inconsistent among the duplicate logging statements in one set. 

For the LM example shown in Table~\ref{tab:patterns}, the common words shared by the first pair (i.e., method {\em doScaleUp()} and its log) are ``scale, up'', while the common word shared by the second pair is ``scale''. Hence, \toolS detects an LM instance due to this inconsistency.
%Note that we ignore the cases where the method only has one word, since we find that such method names are usually generic names such as {\tt run()} and {\tt execute()}.
% \toolS detects a potential instance of LM if the number of common words is inconsistent among the same set of duplicate logs.
The rationale is that the number of common words between the class-method name and the associated logging statement is subject to change if developers make copy-and-paste errors on logging statements (e.g., copy the logging statement in {\em doScaleUp()} to method {\em doScaleDown()}), but forget to update the log message to match with the new method name ``doScaleDown''.
However, the number of common words will remain unchanged (i.e., no inconsistency) if the logging statement (after being pasted at a new location) is updated respectively.

%\phead{Detecting inconsistent log level (IL)}.
%\toolS detects an instance of IL if duplicate logging statements in one set (i.e., have the same static text message) have inconsistent log level. Furthermore, \toolS checks whether an instance of IL belongs to one of the three justifiable cases (IL.1--IL.3) and is justifiable by checking the exception types, if the logging statements are in different branches of the same method, and if developers pass the exception variable to another method in the {\em catch} block. 

\vspace{-0.1cm}
\phead{Detecting duplicate logs in polymorphism (DP)}.
%{\em Detection Approach.}
\toolS generates an object inheritance graph when statically analyzing the Java code. For each overridden method, \toolS checks if there exist any duplicate logging statements in the corresponding method of the sibling and the parent class. If there exist such duplicate logging statements, \toolS detects an instance of DP. Note that, based on the feedback that we received from developers (Section~\ref{sec:manual}), we do not expect developers to fix instances of DP. DP can be viewed more as technical debts~\cite{Kruchten:2012:TDM:2412381.2412847} and our goal is to propose an approach to detect DP to raise the awareness from the research community and developers regarding this issue.

\vspace{-0.1cm}

%% file: texfiles/results.tex
\section{Case Study Results}
\label{sec:results}

\begin{table*}
    \caption{The results of \toolS in RQ1 and RQ2. \label{tab:results}}
    \vspace{-0.3cm}
    \centering
    \resizebox{\textwidth}{!} {
    \tabcolsep=8pt
    \begin{tabular}{c|lccc|ccc|ccc|ccc}

        \toprule
    Research    & & \multicolumn{3}{c|}{\textbf{IC}} & \multicolumn{3}{c|}{\textbf{IE}} & \multicolumn{3}{c|}{\textbf{LM}}  & \multicolumn{3}{c}{\textbf{DP}}\\
    questions   & & Pro. & C.Det. & Det. & Pro. & C.Det. & Det.& Pro. & C.Det. & Det. & Tech. & C.Det. & Det. \\
\midrule
\multirow{6}{*}{
\begin{tabular}[c]{@{}p{3.3cm}@{}} RQ1: How well can DLFinder detect duplicate logging code smells in the five manually studied systems?  \end{tabular}
}
        &\textbf{Cassandra}     & 1 & 1 & 1 & 0 & 0 & 0 & 0  & 0& 4  & 2&  2&  2\\
        &\textbf{CloudStack}    & 8 & 8 & 8 & 4 & 4 & 4 & 27& 24 & 186  &  107& 107& 107\\
        &\textbf{Elasticsearch} & 1 & 1 & 1 & 0 & 0 & 0 & 1  & 0 & 15  & 3 & 3& 3\\
        &\textbf{Flink}         & 0 & 0 & 0 & 2 & 2 & 2 & 4  & 4 & 41  & 24 & 24 & 24\\
        &\textbf{Hadoop}        & 5 & 5 & 5 & 0 & 0 & 0 & 9 & 7 & 44  & 27& 27&  27\\

        &\textbf{Total of RQ1}        & 15 & 15 & 15 & 6 & 6 & 6 & 41 & 35 & 290  & 163& 163&  163\\
        & \textbf{Precision / Recall}     & \multicolumn{3}{c|}{100\% / 100\%} & \multicolumn{3}{c|}{100\% / 100\%} & \multicolumn{3}{c|}{12.1\% / 85.4\%}   & \multicolumn{3}{c}{100\% / 100\%} \\

        %\midrule
        %  & \textbf{Total}         & 15 &15 &15 & 4 & 4  & 4 &37 & 31 & 249 & 0 & 0 & 15 &139 & 139 & 139\\
        \midrule
\multirow{6}{*}{
\begin{tabular}[c]{@{}p{3.3cm}@{}} RQ2: How well can \toolS detect duplicate logging code smells in the additional systems?\end{tabular}
%}& \textbf{Camel}         & 1 & 1 & 1 & 0 & 0 & 0 & 14  & 10 & 95 & N/A & N/A & 3 & 29 & 29 & 29\\
}
       & \textbf{Camel}         & 1 & 1 & 1 & 0 & 0 & 0 & 14  & 10 & 95 & 29 & 29 & 29\\
       & \textbf{Kafka}         & 0 & 0 & 0 & 0 & 0 & 0 & 3  & 3 & 15 & 14 & 14 & 14\\
       & \textbf{Wicket}         & 1 & 1 & 1 & 0 & 0 & 0 & 1  & 1 & 4  & 1 & 1 & 1\\
       &\textbf{Total of RQ2}        & 2 & 2 & 2 & 0 & 0 & 0 &18  & 14 & 114  & 44& 44&  44\\
       & \textbf{Precision / Recall}     & \multicolumn{3}{c|}{100\% / 100\%} & \multicolumn{3}{c|}{- / -} & \multicolumn{3}{c|}{12.3\% / 77.8\%}   & \multicolumn{3}{c}{100\% / 100\%} \\

        \midrule
       %& \textbf{Total}         & 2 &2 & 2 & 0 & 0  & 0 &15 & 11 & 99 & 0 & 0 & 3 &30 & 30 & 30\\
        & \textbf{Total}         & 17 &17 & 17 & 6 & 6 & 6 & 59 &49 & 404 &207 & 207 & 207\\

        % cass non-dup median 7, 0.146 neg
        % cloud stack non-dup median 8, 0.045 neg
        % es non-dup median 7, 0.27 small
        % hadoop non-dup median 6, 0.08 neg
        \bottomrule
    \end{tabular}
    }
        \noindent {\sf Pro.}: number of problematic instances as the ground-truth, {\sf Tech.}: number of technical debt instances for DP, {\sf C.Det.}: the combined number of problematic or technical debt instances {\sf correctly} detected by \tool, {\sf Det.}: number of instances detected by \tool.
    \vspace{-0.6cm}

    \label{table:detection}
\end{table*}

\vspace{-0.1cm}

In this section, we conduct a case study to investigate the prevalence of duplicate logging code smells and evaluate \toolS by answering three research questions.

\phead{RQ1: How well can \toolS detect duplicate logging code smells in the five manually studied systems?}

\noindent{\bf Motivation.}
\toolS was implemented based on the duplicate logging code smells uncovered from the manually studied systems (i.e., IC, IE, LM, and DP). Since we obtain the ground truth (i.e., all the duplicate logging code smell instances) in these five systems from our manual study, the goal of this RQ is to evaluate the detection accuracy of \tool.  

\noindent{\bf Approach.}
We applied \toolS on the same versions of the systems that we used in our manual study (Section~\ref{sec:manual}). We calculated the precision and recall of \toolS in detecting problematic instances for IC, IE, and LM, as well as the technical debt instances for DP. 
Precision is the percentage of correctly detected instances among all the detected instances, and recall is the percentage of problematic or technical debt instances that \toolS is able to detect.

\noindent{\bf Results and discussion.}
The first five rows of Table~\ref{tab:results} show the results of RQ1. For the patterns of IC, IE, and DP, \toolS detects all the problematic and technical debt instances of duplicate logging code smells (100\% in recall) with a precision of 100\%. 
%For the IL pattern, since we do not find any problematic instances (as discussed in Section~\ref{sec:manual}), both of the columns of problematic instances in ground truth ({\em Pro.}) and  correctly detected ({\em C.Det.}) in Table~\ref{tab:results} are 0. 
For the LM pattern, \toolS achieves a recall of 85.4\% (i.e., \toolS detects 35/41 problematic LM instances). We manually investigate the six instances of LM that DLFinder cannot detect. We find that the problem is related to the various habits and coding conventions that developers use when writing log messages. For example, developers may write ``mlockall'' instead of ``mLockAll'' (i.e., the camelcase naming convention), which increases the challenge of log message analysis. Hence, the text in the log message cannot be matched with the method name when we split the word using camel cases. The precision of detecting problematic LM instances is modest because, in many false positive cases, the log messages and class-method names are at different levels of abstraction: The log message describes a local code block while the class-method name describes the functionality of the entire method.
For example, {\em encodePublicKey()} and {\em encodePrivateKey()} both contain the duplicate logging statement {\em ``Unable to create KeyFactory''}. The duplicate logging statement describes a local code block that is related to the usage of the {\em KeyFactory} class, which is different from the major functionalities of the two methods (i.e., as expressed by their class-method names).
Nevertheless, \toolS detects the LM instances with a high recall, and developers could quickly go through the results to identify the true positives (it took the first two authors less than 10 minutes on average to go through the LM result of each system to identify true positives).
%Since we focus on the checking of log message and method/class name rather than the full code context. Although we have tried our best to filter the logging statements which have weak descriptive power for the actual system behavior, there are still a large amount of unexpected cases since the coding habit of different developers vary a lot.

\begin{figure}
    \centering
  \subfloat[Precision\label{comparea}]{%
       \includegraphics[width=0.45\linewidth]{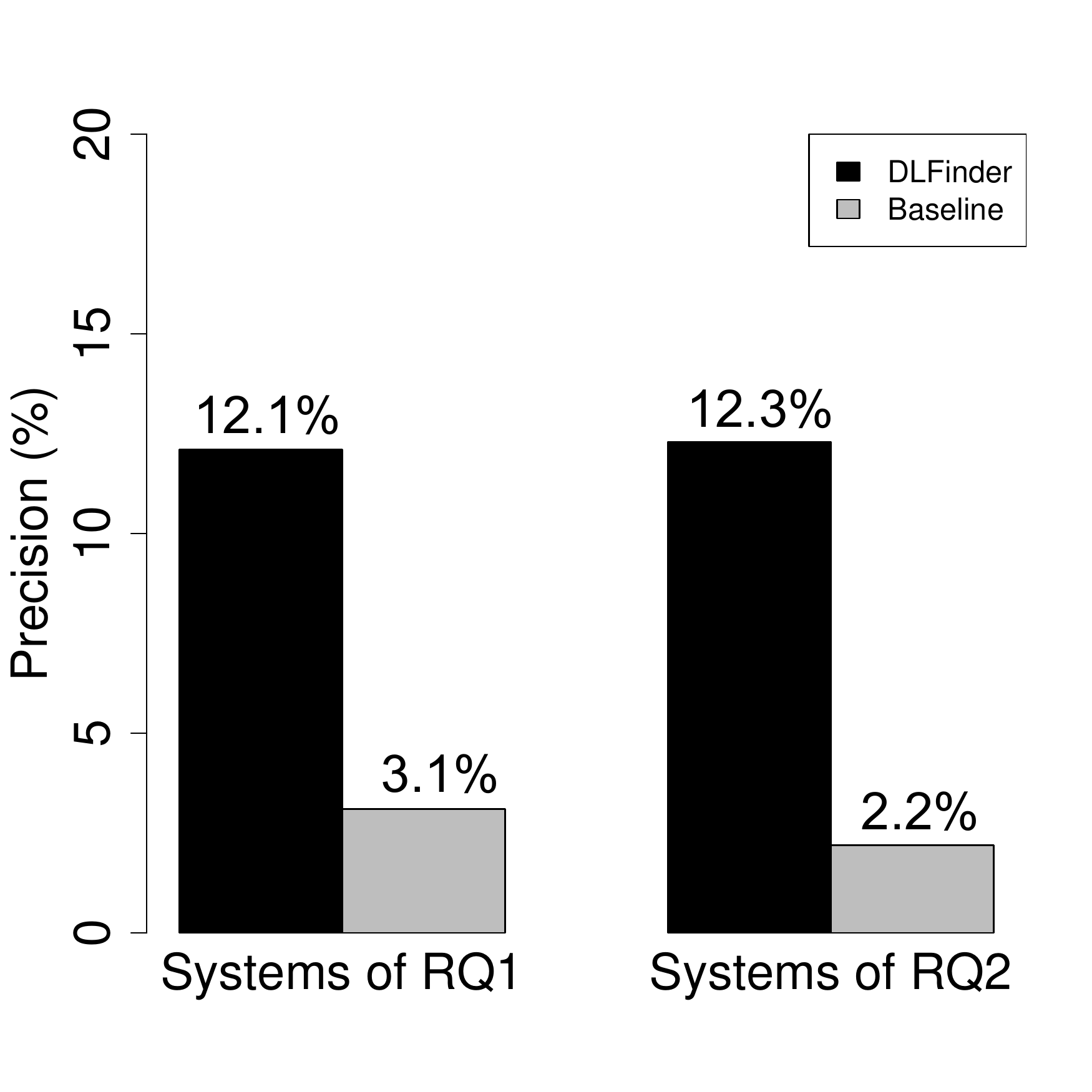}}
    \hfill
  \subfloat[Recall\label{compareb}]{%
        \includegraphics[width=0.45\linewidth]{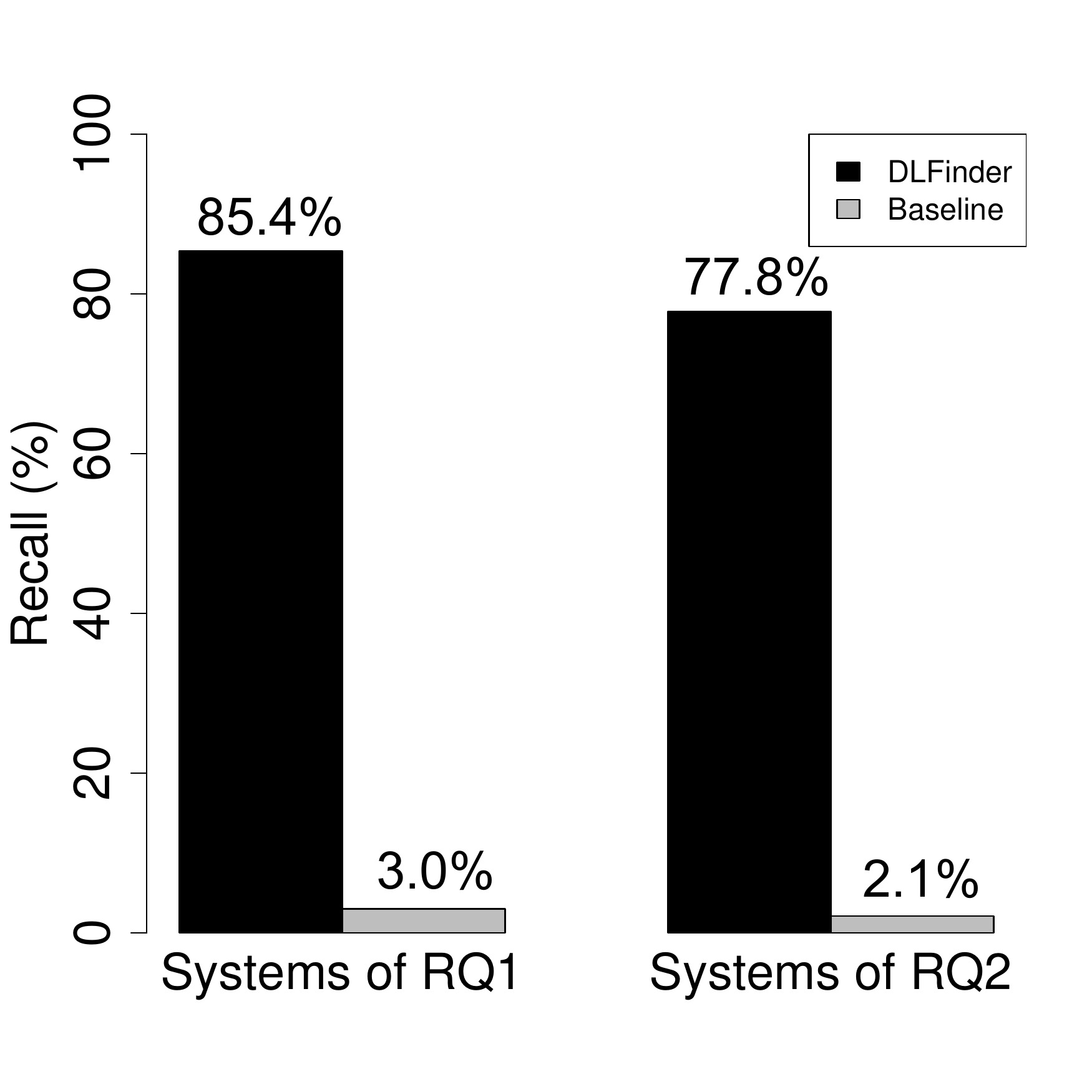}}
    \\

  \caption{The precision (a) and recall (b) of \toolS detecting LM on the systems of RQ1 and RQ2 respectively, compared with the baseline (random prediction).}
  \label{figure:compare}
  \vspace{-0.4cm}
\end{figure}

To further evaluate our detection approach for LM, we compare our detection results with a baseline. We use random prediction algorithm as our baseline, which is commonly used as the baseline in prior studies~\cite{DBLP:conf/esem/XiaSKLW16, ELBlocker, BlockBugs}. The random prediction algorithm predicts the label of an item (i.e., whether a set of duplicate logging statements belong to LM) based on the distribution of the training data. %\peter{Not sure if we need this. This sounds too easy}For example, if 5\% of the duplicate logging statements contain LM, each set of duplicate logging statement has a 5\% chance of being classified as LM. \peter{the result of each system or the combined distribution? what about the distribution in RQ2?}
%\zhenhao{When doing random prediction, the distribution is based on each system. When presenting the results, we combine the results of systems in RQ1 and RQ2 together, respectively.}
For each system, we use our manually labeled results (which are discussed and verified in the previous sections) as the training data. Note that we only compare the detection results of LM with the baseline. The reason is that pattern IC, IE, and DP are relatively independent and well-defined, unlike LM which depends on the semantics of the logging statement and its surrounding code. We repeat the random prediction 30 times (as suggested by previous studies~\cite{30times, peter_icse}) for each system to reduce the biases. Finally, we report the average precision and recall that are computed based on the 30 times of iterations. Figure~\ref{figure:compare} shows how the precision and recall of our approach compared to that of the baseline. The average precision and recall for the baseline are 3.1\% and 3.0\%, respectively, for the five studied systems. Our detection approach achieves a precision and recall of 12.1\% and 85.4\%, respectively. In short, our approach is better than the baseline and is able to have a very high recall in the five manually studied systems.

\phead{RQ2: How well can \toolS detect duplicate logging code smells in the additional systems?}

%We applied \toolS on two additional systems.
%The goal of this RQ is to study whether the uncovered patterns of duplicate logging code smells are generalizable to other systems. We applied \toolS on four additional systems that are not included in the manual study in Section~\ref{sec:manual}: Kafka 2.1.0 (released on Nov. 20, 2018) Flink 1.7.1 (released on Dec. 21, 2018) Camel 2.21.1 (released on Apr. 28, 2018) and Wicket 8.0.0 (released on May 16, 2018), which are all large-scale open source Java systems. Similar to our manual study, the first two authors of this paper manually collect the problematic duplicate logging code smells in the additional systems, i.e., the ground-truth used for calculating the precision and recall of \tool. Note that the collected ground-truth of the additional systems is only used in this evaluation, but not in designing the patterns in \tool. (There are also no new patterns found in this process.)
\noindent{\bf Motivation.}
The goal of this RQ is to study whether the uncovered patterns of duplicate logging code smells are generalizable to other systems.

\noindent{\bf Approach.}
We applied \toolS to three additional systems that are not included in the manual study in Section~\ref{sec:manual}: Camel, Kafka, and Wicket, which are all large-scale open source Java systems. Details of the systems are presented in Table~\ref{table:systems}. Similar to our manual study, the first two authors of this paper manually collect the problematic and technical debt duplicate logging code smells in the additional systems, i.e., the ground-truth used for calculating the precision and recall of \tool. Note that the collected ground-truth of the additional systems is only used in this evaluation, but not in designing the patterns in \toolS (There are also no new patterns found in this process).

\noindent{\bf Results and discussion.}
The second half of Table~\ref{tab:results} shows the results of the additional systems. In total, we found 20 problematic duplicate logging code code smell instances (\toolS detects 16) in these systems and all of them are reported and fixed. Compared to the five systems in RQ1, \toolS has similar precision and recall values in the additional systems.
%\peter{maybe remove this discussion on IL or move/modify it to threat}\toolS detected three instances of IL in Camel; however, based on the manual investigation and developers' feedback, these IL instances are not problematic. Similar to what we discuss in Section~\ref{sec:manual}, the differences in the log level are related to having different semantics in the code. Different from a prior study~\cite{Yuan:2012:CLP:2337223.2337236}, we found that all IL instances are not problematic in the eight evaluated systems. Future studies are needed to investigate the effect of IL. 
\toolS detects DP instances with 100\% in recall and precision; however, 
developers are reluctant to fix them due to limited support from logging frameworks. 
Similar to our observation in RQ1, we find that \toolS cannot detect some LM instances due to the various habits and coding conventions when developers write log messages. We also compare our LM detection results with the baseline mentioned in RQ1 using the same approach. The average precision and recall for \toolS are 12.3\% and 77.8\%, respectively, which are considerably better than the precision (2.2\%) and recall (2.1\%) of the baseline. In summary, apart from the manually studied systems in RQ1, \toolS also achieves noticeably better precision and recall than the baseline and is able to have a reasonably high recall in the additional systems.

%we repeat the random predicion 30 times for each system of RQ2. Everytime we record the number of predictions (P) it makes and number of correct predictions (CP). For each system, we use the average of P and TP to calculate the precision and recall for Random Prediction. The average standard deviation of P in the four manually studies systems is 1.55, of CP is 0.34. As shown in Figure~\ref{figure: compare}, \toolS achieves definitely high recall (above 80\%) in the systems of RQ1 and RQ2, while the recall of baseline is observably low (below 5\%).

\begin{table}
    \caption{\label{RQ3}The results of \toolS in RQ3. 
    %Instances of duplicate logging code smells newly introduced in the newer releases of systems studied in Section\ref{sec:manual} Gap. shows the duration of time (days) between the original release (Org.) studied in Section\ref{sec:manual} and the newer release (New.)
    }
    \vspace{-0.3cm}
    \centering
    \resizebox{\columnwidth}{!} {
    \tabcolsep=8pt

    \begin{tabular}{lll|c|c|c|c}

        \toprule
        & \multicolumn{2}{c|}{\textbf{Releases}} & \multicolumn{1}{c|}{\textbf{IC}} & \multicolumn{1}{c|}{\textbf{IE}} & \multicolumn{1}{c|}{\textbf{LM}} & \multicolumn{1}{c}{\textbf{DP}}\\
        & Org., New. &Gap.  & & & & \\
        \midrule
        \textbf{Cassandra}     & 3.11.1, 3.11.3 & 294 & 0 & 0 & 0 & 1 \\
        \textbf{CloudStack}     & 4.9.3, 4.11.1 & 297 & 5 & 0 & 2 & 0 \\
        \textbf{Elasticsearch}  & 6.0.0, 6.1.3 & 77 & 0 & 0 & 0 & 0 \\
        \textbf{Flink}         & 1.7.1, 1.9.1 & 301 & 0 & 0 & 0 & 1 \\
        \textbf{Hadoop}         & 3.0.0, 3.0.3 & 208 & 0 & 0 & 2 & 21 \\

        \midrule
        \textbf{Total}         & -  & - & 5 & 0 & 4 & 23 \\
        \bottomrule
    \end{tabular}
    }

    Gap.: duration of time in days between the original (Org.) and the newer release (New.).
    \vspace{-0.4cm}
\end{table}

\phead{RQ3: Are new duplicate logging code smell instances introduced over time?}

\noindent{\bf Motivation.}
In this RQ, we investigate if new instances of duplicate logging code smell are introduced during the evolution of systems. An automated detection tool may then help developers detect such problems overtime.

\noindent{\bf Approach.}
We applied \toolS on the latest versions of the five studied systems, i.e., Hadoop, CloudStack, Elasticsearch, Cassandra and Flink, and compare the results with the ones on previous versions. The gaps of days between the manually studied versions and the new versions vary from 77 days to 301 days.

\noindent{\bf Results and discussion.}
Table~\ref{RQ3} shows that new instances of duplicate logging code smells are introduced during software evolution. 
All the detected problematic instances (i.e., instances of IC, IE, and LM) are reported and fixed.
As mentioned in Section~\ref{sec:manual} and \ref{sec:detection}, our goal of detecting DP is to show developers the logging technical debt in their systems. The number of commits for the studied time periods are: 282 commits for Cassandra, 1,097 commits for Cloud Stack, 1,036 for Elasticsearch, 485 commits for Hadoop, and 3,036 commits for Flink. These 9 instances that we detected and fixed were introduced during the studied period. For the systems that we did not find new instances of IC, IE, and LM, the number of commits is either small (e.g., 282 commits for Cassandra) or have fewer log lines (e.g., Elasticsearch has only 1.7K log lines). However, we still find new instances of DP in Cassandra and Flink. In short, we found that duplicate logging code smells are still introduced over time, and an automated approach such as \toolS can help developers avoid duplicate logging code smells as the system evolves.

\rqboxc{The duplicate logging code smells exist in both manually studied and additional systems. In total, \toolS is able to detect 81 out of 91 problematic duplicate logging code smell instances (combining the results of RQ1, RQ2, and RQ3 for pattern IC, IE, and LM). We also find that new instances of logging code smells are introduced as systems evolve.}
%\vspace{-0.2cm}

%the results related to this RQ.
%Since the purpose of this RQ is to show whether new duplicate logging code smells instances are introduced to systems over time, the instances that have already been studied in the previous sections are excluded.
%Our results show that new duplicate logging code smells are introduced during software evolution. \toolS detects 31 new problematic instances in the latest version of the four systems.

%% file: texfiles/rq4.tex
%\section{RQ4: What are the Relationships between Duplicate Logging Statements and Code Clone?}

\section{RQ4: What are the Relationships between Problematic Duplicate Logging Code Smells and Code Clones?}

\label{sec:clone}

\phead{Motivation. }
Code clone or duplicate code is considered a bad programming practice and an indication of deeper maintenance problems~\cite{refactoring1999}. Prior studies often focus on studying clones in source code and understanding their potential impact. However, there may also be other negative side effects that are related to code clones. For example, logging statements can also be copied along with other code since cloning is often performed hastily without much attention on the context~\cite{5463343}. In the previous RQs, we focus on studying problematic and technical debt instances of duplicate logging code smells (i.e., IC, IE, LM, and DP). In this section, we further investigate the potential causes of these instances by examining their relationship with code clones (we refer both the problematic and technical debt instances as problematic instances in this section for simplification). Our findings may provide researchers and practitioners with insights of other possible effects of code clones, other ways to further improve logging practices, and inspire future code clone studies.

\noindent{\bf Our approach of mapping code clones to problematic instances of duplicate logging code smells.}
%We use both an automated and a manual approach to study the relationship between code clones and instances of problematic duplicate logging code smells (i.e., DP and fixed instances of IC, IE, and LM). 
Due to the large number of duplicate logging statements in the studied systems (Appendix~\ref{sec:appendix2} also studies the relationship between general duplicate logging statements and code clone), we first leverage automated clone detection tools to study whether these instances (i.e., DP and fixed instances of IC, IE, and LM) reside in cloned code. In particular, we use NiCad~\cite{nicad} as our clone detection tool. NiCad uses hybrid language-sensitive text comparison to detect clones. We choose NiCad because, as found in prior studies~\cite{nicad,NicadEvaluation}, it has high precision (95\%) and recall (96\%) when detecting near-miss clones (i.e., code clones that are very similar but not exactly the same) and is actively maintained (latest release was in July 2020). Note that, we find NiCad's precision to be 96.8\% in our manual verification, which is consistent with the results from prior studies (more details in Appendix~\ref{sec:appendix2}). Note that, we find NiCad's precision to be 96.8\% in our manual verification, which is consistent with the results from prior studies (more details in Appendix~\ref{sec:appendix2}). 

In NiCad, the source code units of comparison are determined by partitioning the source code into different granularities. %a set of disjoint fragments (e.g., methods or code blocks).
%These source code units are the largest source code fragments (e.g., basic blocks) that are considered in the clone detection~\cite{nicad}. %may be clone targets involved in direct clone relations with each other~\cite{nicad}.
The structural granularity of the source code units could be set as the method-level or block-level (e.g., the blocks of {\em catch}, {\em if}, {\em for}, or {\em method}, etc). In our study, we set the level of granularity to block-level and use the default configuration (i.e., similarity threshold is 70\% and the minimum lines of a comparable code block is 10),
%\jinqiu{Does the paper say the default configurations would yield better results? If so, add that here. Intuitively, 10 sounds too big a number for this configuration though. The example of IL in Table 2 does not have more than 10 lines}
which is suggested by prior studies indicating this configuration could achieve remarkably better results in terms of precision and recall~\cite{EvaluatingModernCloneDetectionTools, ROY2009470, nicad}. Block-level provides finer-grained information, since logging statements are usually contained in code blocks for debugging or error diagnostic purposes~\cite{Fu:2014:DLE:2591062.2591175}. Note that if the block is nested, the inner block is listed twice: once inside its parent block and once on its own. Hence, all blocks with lines of code above the default threshold will be compared for detecting clones. %\peter{I moved the numbers above. We can define what they meant by threshold above as well}For the thresholds and settings of detection, we keep them in default (the minimum lines of a comparable block is 10, the similarity threshold is 70\% \zhenhao{might briefly explain how do they define similarity.})\zhenhao{These are the threshold they used, but didn't describe the reasons in their paper}.
%\peter{avoid passive voice}The study is conducted by answering two research questions. For the identification of cloned code, We use NiCad5 to conduct the detection of code clone. NiCad5 is a flexible hybrid language-sensitive text comparison clone detection tool originally based on the previous work of Roy. {\em et al$.$}~\cite{nicad} with continuously maintainence and improvement afterwards. Up to the submission of this work, the latest version of NiCad5 was released on Oct. 05, 2018. The source unit of comparison are determined by partitioning the source into a set of disjoint fragments. These units are the largest source fragments that may be involved in direct clone relations with each other~\cite{nicad}. The structural granularity of source units could be set as method-level or block-level (e.g., the blocks of {\em catch, if, for, method, etc}). In our study, we set the level of granularity as block-level, a more fine-grained level which might provide us with more accurate results. Note that if the block is nested, the inner block is listed twice: once inside its parent block and once on its own. \peter{Hence, all qualified blocks will be compared by the tool.}%So that all the qualified blocks (with more potential cloned lines than the threshold) could be compared.
We run NiCad on the eight studied open source systems that are mentioned in Section~\ref{sec:results}. We then analyze the clone detection results and match the location of the clones with that of problematic instances. If two or more cloned code snippets contain the same set of instances, we consider the instances are related to the clone. 

To reduce the effect of false negatives, we also manually study the code of {\em all} the remaining instances that are not identified as clones by NiCad. We manually classify the clones into the three following categories: %If the missed clones reside in code blocks that have fewer than 10 lines of code, we manually classify them as micro-clones. 

    %\item Clones (C): if at least two of the logging statements in the set meet the following requirements: 1) The surrounding code block that the logging statement locates in has 5 or more lines of code. \zhenhao{the reason is, it's hard to say if short blocks are clones or just by chance.} 2) At least 5 lines of code surrounding the logging statement are exactly the same, or nearly the same, only with the differences of identifier names.
{\em Clones}: The code around the logging statements is more than 10 lines of code (same as the threshold of the clone detection tool). The code is exactly the same, or only with differences in identifier names (i.e., Type 1 and Type 2 clones~\cite{CompareCloneDetectionTools2007}) but not detected by the clone detection tool.

{\em Micro-clones}: The code around the logging statements is very similar but is less than the minimum size of regular code clones~\cite{microclones}. Prior studies show that micro-clones are also important for consistent updates and they are more difficult to detect due to their small size~\cite{microclone4,microclone5,microclones}. However, the effect of micro-clones on code maintenance and quality is similar to regular code clones~\cite{MicroclonesAndBugsICPC, MicroclonesAndBugsSANER}. Micro-clones should not be ignored when making decisions of clone management.
     %1) The code block where the logging code statements are located is short (i.e., has fewer than ten lines of code\peter{might need to remove the rest?}, below the threshold of the tool). 2) The code in the block is exactly the same, or only with differences in identifier names. For this kind of situations, though we feel they are likely to be clones, their sizes are not long enough for us to make a decision. Therefore, we label them as "Undecidable".
     %Prior studies indicate that micro-clones have similar tendencies of replicating severe bugs as regular clones~\cite{MicroclonesAndBugsICPC, MicroclonesAndBugsSANER}.

{\em Non-clones}: We classify other situations as non-clones.

\begin{table*}
  \caption{The results of code clone analysis on problematic instances and code clones. } 
    \vspace{-0.3cm}
    \centering
    \resizebox{\textwidth}{!} {
    \scalebox{0.8}{
    \tabcolsep=2pt
    \renewcommand\arraystretch{1.1}
    \begin{tabular}{c|rrrr|rrrr|rrrr|rrrr}

        \hline
    & \multicolumn{4}{c|}{\textbf{IC}} & \multicolumn{4}{c|}{\textbf{IE}} & \multicolumn{4}{c|}{\textbf{LM}} & \multicolumn{4}{c}{\textbf{DP}}\\
    & Clone (A) & Clone (M) &Micro.& Clone/Total & Clone (A) & Clone (M)  & Micro.& Clone/Total & Clone (A) &Clone (M) & Micro.& Clone/Total  & Clone (A)  &Clone (M)   &Micro. & Clone/Total\\
\midrule

        \textbf{Cassandra}     & 0& 0 & 0  & 0/1     & 0  &0 & 0 & 0/0       &0 & 0&  0  &0/0     & 2&  0  &0 &2/2\\
        \textbf{CloudStack}     & 5  &0 & 3 & 8/8    &4 & 0& 0 & 4/4       &  20 & 1  &5  & 26/27     & 60&  12  &9&81/107\\
        \textbf{Elasticsearch}  & 0   &0 & 0 & 0/1    &0   &0 & 0  & 0/0      &0  & 0 &1 &1/1    & 0&  1  &0 & 1/3\\
        \textbf{Flink}          & 0  &0 & 0 & 0/0     & 1 & 0  & 1 & 2/2     & 2 & 0 &2 &4/4     & 19&  0 &3 &22/24\\
        \textbf{Hadoop}         & 1  &0 & 2 & 3/5    &0   &0& 0 & 0/0        & 0& 3  &3 & 6/9     & 5&  14  &6 & 25/27\\

       \textbf{Camel}          & 0  &0 & 1 & 1/1    & 0  & 0 & 0 & 0/0       & 6 & 2 &6&14/14     & 22&  2  &4 & 28/29\\
       \textbf{Kafka}          & 0   &0 & 0  & 0/0    & 0  &0 & 0  & 0/0       & 1 & 0  &1 &2/3     & 3&  3  &2 & 8/14\\
       \textbf{Wicket}         & 0  &0 & 0  & 0/1   & 0  &0 & 0  & 0/0       & 0 & 1  &0 &1/1     & 1&  0  &0 & 1/1\\

        \hline

        \textbf{Total}         &6  & 0 & 6  & 12/17   &5 & 0 &1&6/6      &29 & 7 &18 &54/59     &112 & 32 &24 &168/207\\

        % cass non-dup median 7, 0.146 neg
        % cloud stack non-dup median 8, 0.045 neg
        % es non-dup median 7, 0.27 small
        % hadoop non-dup median 6, 0.08 neg
        \hline
    \end{tabular}
    }
    }
     \noindent {\sf Clone (A)}: number of problematic duplicate logging code smell instances that are detected as clones by NiCad, {\sf Clone (M)}: number of problematic duplicate logging code smell instances that are identified as clones by manual study, {\sf Micro.}: number of problematic duplicate logging code smell instances that are identified as Micro clones by manual study.

    \vspace{-0.6cm}

    \label{table:RQ5}
\end{table*}

\noindent{\bf Result of code clone analysis on problematic instances.}
{\em We find that 240 out of 289 (83\%) of the problematic instances of duplicate logging code smells reside in cloned code snippets.}
Table~\ref{table:RQ5} presents the results of our  code clone analysis. {\sf Clone (A)} refers to the number of problematic instances that are detected by NiCad as in code clones. {\sf Clone (M)} refers to the number of problematic instances that are manually found as in code clones. {\sf Micro.} refers to the number of problematic instances that are manually found as in Micro clones (i.e., less than 10 lines of code). In general, our findings show that these problematic instances are potentially caused by code clones. {\bf In other words, in addition to the finding from prior code clone studies, which indicates that code clones may introduce subtle program errors~\cite{contextCloneBugs, tracyhallcodesmell}, we find that code clones may also result in bad logging practices that could increase maintenance difficulties.} Future studies should further investigate the negative effect of code clones on the quality of logging statements and provide a comprehensive logging guideline.%of duplicate logging code smells are closely related to code clones

{\em We find that 64.2\% (88/137) of the problematic instances of duplicate logging code smells that are labeled as Non-clones by the automated code clone detection tool are actually from cloned code snippets. Among them, more than half (55.7\%, 49/88) reside in micro clones, which often do not get enough attention in the process of code clone management.}
As mentioned in the approach section of this RQ, to overcome potential false negatives, we manually study {\em all} the 137 problematic instances that are labeled as Non-clones by NiCad. %Table~\ref{table:RQ5} presents the results of our manual study for each pattern. {\sf Clone (M)} refers to the number of problematic instances that are identified our manual study as in code clones. {\sf Micro. } refers to the number of problematic instances that are identified as in Micro clones by our manual study. 
We classify each instance that we study into three categories: 

{\em \underline{Category 1:} Code clones reside in part of a large code block.} Since the structural granularity level of the source code units is block-level (i.e., the minimal comparable source code unit of the tool is a block), the similarity of the code is computed by comparing blocks. However, developers may copy a small part of the code into a large code block. In such cases, the similarity would be low between two different large code blocks which only have a few lines of cloned code.  %For example, developers may copy the code (along with the logging code statements) that is related to establishing network connect from one file to another. However, the copied code is just part of another complex method; hence, the clone detection tool fails to detect the code snippets as clones since the minimum detectable source code unit is block.

{\em \underline{Category 2:} Code clones reside in code with very similar semantics but have minor differences.} The surrounding code of duplicate logging statements share highly similar semantics (i.e., implement a similar functionality), but have minor differences (e.g., additions, deletions, or partial modification on existing lines). Such scattered modifications might reduce the similarity between the code structures, and thus, result in miss detection~\cite{ROY2009470,nicad}. For example, there is a code block in {\tt\small \textbf{FTP}Consumer} of Camel which does a series of operations based on the file transfer protocol (FTP). %The code in the block may be reuseable for other similar ptopocols. Thus,
Due to the similarity between FTP and secured file transfer protocol (SFTP), Camel developers copied the code block and made modifications (e.g., change class and method names) to the all the places where SFTP is needed (e.g., {\tt\small \textbf{SFTP}Consumer}). Therefore, clone detection tools may fail to detect this kind of cloned code blocks as due to minor yet scattered changes.

{\em \underline{Category 3:} Short methods/blocks.}
The logging statements reside in very short methods or code blocks with only a few lines of code. For example, there is a method in CloudStack named {\em verifyServicesCombination()} containing only six lines of code and duplicately locates in three different classes. The method verifies the connectivity of services, and generates a warning-level log if it fails the verification. Clone detection tool fails to detect this category of cases due to their small size compared to regular methods.

\noindent{\bf {\em IC \& IE: }} {\em 30\% (7/23) of the IC and IE instances in cloned code are related to micro-clones.} Since both of IC and IE reside in {\em catch} blocks, which usually contain only a few lines of code, we discuss these two duplicate logging code smells together. As shown in Table~\ref{table:RQ5}, , 7 (6 IC + 1 IE) out of 23 (17 IC + 6 IE) instances are labeled as {\em Micro-clones}, and 11 instances are identified as clones by the clone detection tool. The remaining five instances are labeled as {\em Non-clones}, since they are single logging statement thrown with multiple types of exceptions (e.g., {\sf catch (Exception1 \(|\)  Exception2 e)}). We find that all of the seven {\em Micro-clones} instances belong to Category 1 (i.e., short code snippets within a large code block). The reason might be that these logging statements all reside in {\em catch} blocks, which are usually very short. Thus, although the code in these short code blocks are identical or highly similar, they are not long enough to be considered as comparable code blocks by the clone detection tool. %Their parent blocks may have sufficient length, but the cloned code only occupies a small portion of their parent blocks. Thus, the tool fails to detect them due to the insignificant similarity of their parent blocks.

\noindent{\bf {\em LM: }}{\em 25/54 (46\%) of the LM instances in cloned code cannot be detected by automated clone detection tools. 92\% (54/59) of LM are related code clones.} As shown in Table~\ref{table:RQ5}, 36 out of 59 instances are labeled as {\em Clones} (29 instances by tool + 7 instances by manual study), 18 out of 59 instances are labeled as {\em Micro-clones}, and the remaining 5 instances are labeled as {\em Non-clones}. For the seven instances that are identified as {\em Clones} by manual study, they all belong to Category 2 (i.e., they share highly similar semantics, but have minor differences). The reason might be that developers copy and paste a piece of code along with the logging statement to another location, and apply some modifications to the code. However, developers forgot to change the log message. Similarly, for the five instances that are labeled as {\em Non-clones}, we find that even though the code is syntactically different, the log messages do not reflect the associated method. For the 18 {\em Micro-clones} instances, 11 out of 18 instances belong to Category 3 (short methods), and the remaining 7 are Category 2 (short code snippets within a larger code block). As confirmed by the developers (in Section~\ref{sec:manual}), these LM instances are related to logging statements being copied from other places in the code without the needed modification (e.g., updating the method name in the log). 

Our manual analysis on LM instances provides insights on possible maintenance problems that are related to the modification and evolution of cloned code. Moreover, 92\% of the LM instances are related to code clones. Future studies may further investigate the inconsistencies in the source code and other software artifacts (e.g., logs or comments) that are caused by code clone evolution.

\noindent{\bf {\em DP: }}{\em 81\% (168/207) of the DP instances are either Clones or Micro-clones, which shows that developers may often copy code along with the logging statements across sibling classes.} In total, 144 out of 207 DP instances are labeled as {\em Clones} (112 by tool + 32 by manual study), 24 are labeled as {\em Micro-clones}, and the remaining 39 instances are {\em Non-clones}. For the 32 instances that are labeled as {\em Clones} by manual study, 16 instances are Category 1 (part of a large code block), the remaining 16 instances are Category 2 (very similar semantics with minor differences). For the 24 {\em Micro-clones} instances, 11 instances belong to Category 3 (short methods), and the remaining 13 are categorized as Category 1 (short code snippets within a larger code block).
%Combined with the results from the clone detection tool, around 70\%$\sim$81\% (\peter{the sentence in the bracket is not clear}144/207~\jinqiu{Where is this 207 from?} without any instances of Undecidable, 167/207 with all the instances of Undecidable \zhenhao{(112+32)=144 Clones, with 24 Undecidable instances out of 207 total instances. If 0 Undecidable instance is considered, then 70\%. If all, then 81\%. }) of the DP instances are related to code clones.
Combined with the results from the clone detection tool, 81\% (112 detected by the tool + 32 Clones + 24 Micro-clones identified by manual study, out of 207 total instances) of the DP instances are related to code clones.
One possible reason that many DP instances are related to code clone is that DP is related to inheritance. Classes that inherit from the same parent class may share certain implementation details. Nevertheless, due to the similarity of the code, developers should consider updating the log messages to distinguish the executed methods during production to assist debugging runtime errors.

For all of the remaining problematic instances (49/289) that are not classified as clones by the automated tool and manual analysis, they mostly reside in very short code blocks (e.g., only 1$\sim$3 lines of code). Even though these code blocks may be similar or even identical, we cannot tell whether they are clones or not. It is possible that developers implemented such similar code by coincidence, or the code was copied from other places and are then modified (but forgot to modify the log-related code).

%However, as discussed in Section~\ref{sec:manual}, resolving DP often requires systematic refactoring, and the fixes of DP instances are not definitely expected as the technical debts need to be considered~\cite{Kruchten:2012:TDM:2412381.2412847}. \zhenhao{what can we learn from the view of code clones?}

%\jinqiu{I am forming one possible highlight here, not necessarily a correct argument: Our finding indicates that semantic information such as duplicate logging statements may serve as trails to distill copy-and-paste clones from general clones, which are commonly considered as bad programming practices and a source of bugs.}

%In summary, our finding indicates that the semantic information in the code, such as logging statements, may be used as trails to help

%\rqbox{Take-home box. \zhenhao{Todo: Talk about sth like LM instances usually locate in short blocks (shorter than 10 lines), so it might be hard for the existing clone detection approaches to find this kind of problems at the meantime of addressing code clone problems. More specific approaches or tools are needed to tackle it.}
%\zhenhao{Still a large portion of problematic instances reside in short methods / short blocks, current clone detection tools might not be able to detect them effectively.}\peter{@Zhenhao, see if you can come up with the takehome based on the updated discussions}
%}
%\vspace{-0.1cm}

\noindent{\bf Implication and highlights of our code clone analysis.} Our finding shows that most of problematic instances of duplicate logging code smells are indeed related to code clones, and many of which cannot be easily detected by state-of-the-art clone detection tools. Our finding shows additional maintenance challenges that may be introduced by code clones -- maintaining logging statements and understanding the runtime behaviour of system execution. %We find that duplicate logging statements may also be a possible indication of hard-to-detect clones based on our manual study. Moreover, most of the problematic instances of duplicate logging code smell reside in cloned code snippets. 
Hence, future code clone detection studies should consider other possible side effects of code clones in addition to code maintenance and refactoring overheads. Future studies may also consider integrating different information in the software artifacts (e.g., duplicate logging statements or comments) to further improve clone detection results. %when implementing clone detection tools. %that for the code snippets that are not detected as clones

\rqboxc{83\% of the problematic instances of duplicate logging code smells (240 out of 289 instances, combining the results of tool detection and manual study) are related to code clones. Our finding further shows the potential negative effect of code clones on system maintenance. Moreover, 17\% of the instances reside in short code blocks, which might be difficult to detect by using existing code clone detection tools.
}
%\vspace{1mm}
\noindent{\bf Discussion: the potential of using code clone detection tool to assist in finding problematic instances of duplicate logging code smells.} In the previous section, we found that most of the problematic instances of duplicate logging code smells (83\%) are related to code clones. Therefore, we use the results of our code clone analysis to compare with and/or assist our detection approach of LM. We focus on studying LM for two reasons. First, we found that 92\% (54/59) of the LM instances are related to code clones. Second, unlike other patterns that have a detection accuracy of 100\%, our current detection approach for LM analyzes textual similarity of the logging statement and its surrounding code, which has a lower precision and recall. Using clone detection results may further help improve our detection accuracy. 

\vspace{-0.05cm}
We first use the clone detection result as a baseline and compare the results with the detection approach implemented in \tool. If two duplicate logging statements reside in cloned code, we consider them as a possible instance of LM. Overall, the average precision and recall of using clone detection result are 3.7\% and 53.7\%, respectively, in the studied systems in RQ1. The average precision and recall in the additional systems in RQ2 are 1.5\% and 38.9\%, respectively. Compared to using clone detection result as a baseline, our approach has a better precision and recall (around 12\% in precision and 80\% in recall). However, among the 10 LM instances that cannot be detected using our approach, four of them are detected by this baseline approach. After manual investigation on these four instances, we found the log message describes a local code block while the class-method name describes the functionality of the entire method. Hence, in such cases, using clone detection results may be more effective in detecting LM.

Inspired by the analysis result, we then study if clone detection result can assist \toolS in finding LM. We use the automated clone detection results from NiCad to filter the LM instances that are detected by \tool. Namely, \toolS only reports that a set of duplicate logging statements is a potential LM instance if they reside in cloned code. We find that, after using clone detection results to filter out potential false positives, the average precision and recall for the eight studied systems are 17.7\% and 42.4\%, respectively. Compared to \tool's detection result (Table~\ref{table:detection}), the precision increases by around 5\% but the recall decreases by around 40\%. The reason may be that many problematic LM instances reside in code clones that are difficult to detect by clone detection tool (e.g., micro clones). As shown in Table~\ref{table:RQ5}, NiCad only detects 29/54 of the LM instances that reside in cloned code. As we discussed in Section~\ref{sec:results}, we believe that recall is more important when detecting LM, since we found the manual effort of evaluating LM instances to be small (i.e. within a few minutes). Our findings also shed light on balancing the precision and recall of detecting duplicate logging code smells. Future studies may consider further improving code clone detection techniques to detect code smells that are related to logging statements. %Future code clone studies may consider improving the detection of micro clones %Nevertheless, our findings also show that, with better clone detection results, 

%% file: texfiles/rq5.tex
\section{RQ5: What are the Relationships Between Duplicate Logging Statements and Code Clones?}
\label{sec:rq5}

\phead{Motivation. }
In Section~\ref{sec:clone}, we investigate the relationship between code clones and {\em problematic instances of duplicate logging code smells}. As discussed in Section~\ref{sec:manual}, duplicate logging code smells are duplicate logging statements with specific patterns that may be indications of logging problems. In this section, we further investigate the relationship between {\em duplicate logging statements} and code clones. We also study the potential impact of duplicate logging statements on detecting code clones.

\phead{Approach. }
Similar to Section~\ref{sec:clone}, we use both an automated and a manual approach to study the relationship between code clones and duplicate logging statements. We first leverage NiCad to automatically detect clones. Although we found that NiCad has a great precision (i.e., 96.8\%, as shown in Appendix~\ref{sec:appendix2}), there may still exist false negatives (i.e., the duplicate logging statements are code clones, but are missed by the tool). Therefore, we manually investigate a statistical sample of duplicate logging statements, which reside in code snippets that are classified by NiCad as {\em Non-clones} to study the false negative rate.

\noindent{\bf Results of automated code clone analysis on duplicate logging statements.} {\em We find that a considerable number of duplicate logging statements (43.7\% on average) reside in cloned code snippets.}
Table~\ref{table:RQ4} presents the results of our code clone analysis. {\sf DupSet} refers to the total sets of duplicate logging statements (a set contains two or more logging statements with the same text message). {\sf CloneSet} refers to the subset of duplicate logging statement sets ({\sf DupSet}) that are from cloned code snippets. The percentage number is the proportion of {\sf CloneSet} out of {\sf DupSet}. Finally, {\sf Avg. Sim.} refers to the average code clone similarity score among the cloned code snippets. As shown in Table~\ref{table:RQ4}, 11.5\% to 51.1\% sets of duplicate logging statements are from the cloned code snippets in the studied systems. Overall, 1,042 out of 2,382 (43.7\%) sets of duplicate logging statements are related to code clones (with an average 80\% similarity score).

Our finding shows that a considerable number of duplicate logging statements are related to code clones, and developers may not change the log messages when they copy a piece of code to another location.
However, due to the importance of logging for understanding system runtime behaviour~\cite{petericseseip2017, Yuan:2012:CLP:2337223.2337236, Yuan:2011:ISD:1950365.1950369}, developers should avoid directly copying logging statements. Developers should consider modifying the log messages (e.g., to include the class name, modify the message to reflect code changes, or record new important dynamic variables) to assist debugging and workload understanding.

\noindent{\bf Results of manual code clone analysis on duplicate logging statements.} {\em We find that more than 50\% of the sampled duplicate logging statements reside in cloned code snippets that are difficult to detect using automated code clone detection tools. In particular, 24.5\% of the manually studied duplicate logging statements are related to code clones, and 26.2\% are related to micro-clones. }
In total, we randomly sample 298 sets of duplicate logging statements to achieve a confidence of 95\% and a confidence interval of 5\%. For each set of the sampled duplicate logging statements, we manually classify them into three types: {\em Clone} (i.e., but not detected by code clone detection tools), {\em Micro-clone} (i.e., code blocks with less than 10 lines of code), and {\em Non-clone}.

Table~\ref{table:manualrq4} presents the results of our manual study. Overall, 73 out of the 298 (24.5\%) manually-studied sets of duplicate logging statements are labeled as {\em Clones}. 78 out of 298 (26.2\%) sets are labeled as {\em Micro-clones}. The remaining 147 out of 298 (49.3\%) sets are labeled as {\em Non-clones}. For 42 out of the 73 cases of {\em Clones}, and 32 out of 78 cases of {\em Micro-clones}, we find that developers often only copy and paste part of the code into another large code block (Category 1 discussed in Section~\ref{sec:clone}). Hence, only small parts of large code blocks are similar, which reduces the similarity score. For 53 out of the 73 cases that are manually identified as {\em Clones}, they reside in code with very similar semantics but have minor differences (Category 2). Note that some cases belong to multiple categories. For 46 out of 78 cases they are classified as {\em Micro-clones}, which reside in very short methods with only a few lines of code (Category 3).

In summary, we find that more than half of the duplicate logging statements reside in cloned code snippets. Our manual study also highlights that many duplicate logging statements reside in cloned code that may be difficult to detect by clone detection tools.

\begin{table}
    \caption{Automated code clone analysis results on duplicate logging statements.}

    \vspace{-0.3cm}
    \centering
    \resizebox{\columnwidth}{!} {
    \tabcolsep=18pt
    \scalebox{0.8}{

    \begin{tabular}{l|c|c|c}

        \toprule
        & \multicolumn{1}{c|}{\textbf{DupSet}} & \multicolumn{1}{c|}{\textbf{CloneSet}}  & \multicolumn{1}{c}{\textbf{Avg. Sim.}} \\

        \midrule
        \textbf{Cassandra}     & 46  & 14 (30.4\%) & 79.7  \\
        \textbf{CloudStack}     & 865  & 442 (51.1\%) & 80.3  \\
        \textbf{Elasticsearch}  & 40  & 17 (42.5\%) & 72.2  \\
        \textbf{Flink}         & 203  & 92 (45.3\%) & 78.8 \\
        \textbf{Hadoop}         & 217  & 25 (11.5\%) & 76  \\

        \textbf{Camel}         & 886  & 421 (47.5\%) & 80.7  \\
        \textbf{Kafka}         & 104  & 23 (22.1\%) & 75.4  \\
        \textbf{Wicket}         & 21  & 8 (38.1\%) & 83.1 \\
\midrule
        \textbf{Overall}         & 2,382   & 1,042 (43.7\%) & 80.0  \\
        \bottomrule
    \end{tabular}
    }
    }
    \noindent {\sf DupSet}: Total sets of duplicate logging statements, {\sf CloneSet}: Sets of duplicate logging statements that are from cloned code snippets,  {\sf Avg. Sim.}: Average similarity of the cloned code snippets.
    \vspace{-0.3cm}
\label{table:RQ4}
\end{table}

\begin{table}
  \caption{Manual study results on the recall of clone detection tool on duplicate logging statements. Both the {\sf Clones} and {\sf Micro-clones} are labeled manually and they are not detected by the clone detection tool. }
    \vspace{-0.3cm}
    \centering
    %\resizebox{\textwidth}{!} {
    \tabcolsep=8pt
    \scalebox{0.9}{
    \begin{tabular}{l|c|c|c|c}

        \toprule
     &{\textbf{Clones}} &{\textbf{Micro-clones}} &{\textbf{Non-clones}} &{\textbf{Total}}  \\
\midrule

        \textbf{Cassandra}    & 1 & 3 & 3 & 7  \\
        \textbf{CloudStack}    & 22 & 26 &46  & 94 \\
        \textbf{Elasticsearch} & 1 & 1 & 3 & 5 \\
        \textbf{Flink}       & 5 & 4 & 16 & 25\\
        \textbf{Hadoop}      & 12 & 6 & 25  & 43 \\

       \textbf{Camel}       & 28 & 30 & 45 & 103 \\
       \textbf{Kafka}       & 3 & 7 & 8 & 18 \\
       \textbf{Wicket}       & 1 & 1 & 1 & 3 \\

        \midrule

        \textbf{Total}       & 73 &78 & 147 & 298 \\

        \bottomrule
    \end{tabular}}
    %}
    \vspace{-0.3cm}

    \label{table:manualrq4}
\end{table}

\noindent{\bf Discussion: The Potential Impact of Duplicate Logging Statements on Detecting Code Clones.}
\label{sec:appendix3}
In this RQ, we find that a noticeable number of duplicate logging statements reside in cloned code snippets. %In order to seek whether duplicate logging statements helps detectthe potential of improving code clone detection using duplicate logging statements, 
We further investigate the impact of duplicate logging statements on the detection of code clones, namely, whether considering duplicate logging statements helps detect code clones. Specifically, for each set of 
{\sf CloneSet} presented in Table~\ref{table:RQ4}, we first remove the duplicate logging statements from the related code snippets. We then re-examine how many code snippets related to prior {\sf DupSet} are still identified as cloned code snippets and how many are not, by using NiCad.

Table~\ref{table:appendixc} shows the results of our experiments on investigating the impact of duplicate logging statements on detecting code clones. {\sf CloneSet} refers to the sets of cloned code snippets with duplicate logging statements. {\sf CloneSet-NDL} refers to the sets of cloned code snippets after removing the related duplicate logging statements. {\sf CloneSet-Reduced} represents the number of sets reduced by comparing {\sf CloneSet-DL} with {\sf CloneSet-NDL}. {\sf Per. Reduced} shows the percentage of {\sf CloneSet-Reduced} given {\sf CloneSet-DL}. On average, 28.4\% {\sf CloneSet} are not detected by NiCad as cloned code snippets after removing duplicate logging statements. Specifically for each studied system, the reduction ranges from 25.0\% in Wicket to around a 47.1\% in Elasticsearch.

We then manually investigate the code snippets that are not detected as cloned code snippets after removing duplicate logging statements (i.e., {\sf CloneSet-Reduced}). We find two potential reasons that the clone detection tool could not detect them as cloned code snippets. 1) {\em Reduced total lines of similar code after removing duplicate logging statements:} The logging statements usually span across one to three, and sometimes even more, lines of code. However, these lines of code in the duplicate logging statements are the main part of the clones. After removing the duplicate logging statements, the total number of similar lines of code snippets is too small for a clone detection tool to consider as clones. 2) {\em Reduced similarity after removing duplicate logging statements:} Duplicate logging statements have exactly the same log message and are represented as Method Invocation nodes in the Abstract Syntax Tree. Removing duplicate logging statements will decrease the similarity of code snippets, both syntactically and semantically. Hence, the similarity might become smaller than the threshold of the clone detection tool and the code snippets are not detected as clones.

In summary, we find that a large portion of the cloned code snippets with duplicate logging statements (from 25.0\% to 47.1\%) are not detected as cloned code snippets after removing the duplicate logging statements. The results show that duplicate logging statements have a non-negligible impact on the detection of code clones. Future code clone studies may consider the effect of logging code in order to further improve the code clone detection techniques.

\rqboxc{More than half of the duplicate logging statements reside in cloned code snippets, and a large portion of them reside in short code blocks which are difficult to detect using existing code clone detection tools. We also find that duplicate logging statements have a non-negligible impact on helping the detection of code clones. Future works may leverage duplicate logging statements to further improve code clone detection tools.
}

\begin{table}
  \caption{The results of investigating the impact of duplicate logging statements on detecting code clones. }
    \vspace{-0.2cm}
    \centering
    \resizebox{\columnwidth}{!} {
    \tabcolsep=5pt
    \renewcommand\arraystretch{1.1}
    \scalebox{1.0}{
    \begin{tabular}{l|c|c|c|c}

        \toprule
     &{\textbf{CloneSet}} &{\textbf{CloneSet-NDL}} &{\textbf{CloneSet-Reduced}} &{\textbf{Per. Reduced }}  \\
\midrule

        \textbf{Cassandra}    & 14 & 10 & 4 & 28.6\%  \\
        \textbf{CloudStack}    & 442 & 329 & 113  & 25.6\% \\
        \textbf{Elasticsearch} & 17 & 9 & 8 & 47.1\% \\
        \textbf{Flink}       & 92 & 64 & 28 & 30.4\%\\
        \textbf{Hadoop}      & 25 & 16 & 9  & 36.0\% \\

       \textbf{Camel}       & 421 & 299 & 122 & 29.0\% \\
       \textbf{Kafka}       & 23 & 13 & 10 & 43.5\% \\
       \textbf{Wicket}       & 8 & 6 & 2 & 25.0\% \\

        \midrule

        \textbf{Total}       & 1042 &746 & 296 & 28.4\% \\

        \bottomrule
    \end{tabular}}
    }
    \vspace{-0.5cm}

    \label{table:appendixc}
\end{table}

%% file: texfiles/threats.tex
\section{Threats to Validity}
\label{sec:threats}
\vspace{-0.1cm}

\phead{Construct validity.}
In this paper, we study duplicate logging statements from a static point of view. There may be other types of unclear log messages that are dynamically generated during system runtime. Using such dynamic information can also be helpful in identifying unclear log messages. However, the generated log messages are highly dependent on the executed workloads (i.e., hard to achieve a high recall). \toolS statically identifies and improves duplicate logging statements, is useful as it does not require any run-time information. Future studies may consider studying runtime-generated logs and further improve logging practices. 
We detect duplicate logging code smells by analyzing the surrounding code of logging statements as their context. Apart from that, the sequence of generated logs may also provide context information (e.g., the relationship among preceding logs and subsequent logs). However, for most of the duplicate logging code smells discussed in this paper, they are not directly related to the log sequences (e.g., the patterns of IC and IE are related to the logging statements and their surrounding catch blocks). Even though analyzing the generated log sequences may provide more information, the duplicate logging code smells can still cause challenges and increase maintenance costs, as acknowledged by the developers in the studied systems. Future study may consider the execution path of logging statements as the context information to further improve logging practice.

\phead{Internal validity.}
We conducted manual studies to uncover the patterns of duplicate logging code smells, study their potential impact and examine duplicate logging statements that are not classified by the automated clone detection tool as clones. Involving external logging experts may uncover more patterns of logging statements or have different manual study results. To mitigate the biases, two of the authors examine the data independently. For most of the cases the two authors reach an agreement. Any disagreement is discussed until a consensus is reached. In order to reduce the subjective bias from the authors, we have contacted the developers to confirm the uncovered patterns and their impact.
When detecting LM instances, using different approaches to split the text into words may have different results. We follow common text pre-processing techniques to split the text by space and camel case~\cite{Chen:2016:SUT:2992358.2992444}.
We define duplicate logging statements as two or more logging statements that have the same static text message. We were able to uncover five patterns of duplicate logging code smells and detect many duplicate logging code smell instances. However, logging statements with non-identical but similar static texts may also cause problems to developers (e.g., when analyzing dynamically generated logs). Future studies should consider different types of duplicate logging statements (e.g., logs with similar text messages). 
We remove the top 50 most frequent words when detecting LM, because there is a considerable number of generic words across different log messages. However, this might also introduce false negatives. Future studies may consider applying more advanced techniques to better detect the instances of LM. 
There is a considerable number of code clone detection tools proposed by prior studies~\cite{kamiya2002,cpminer,DCCFINDER,duplix,nicad,gabel}. We use NiCad~\cite{nicad} to detect code clone, as it has high precision (95\%), recall (96\%) and outperforms the state-of-the-art code clone detection tools~\cite{nicad,NicadEvaluation,ROY2009470} when detecting near-miss clones, and is actively maintained (latest release was in July 2020). We also manually examine the precision of Nicad in Appendix~\ref{sec:appendix2}, where we find its precision to be 96.8\% in our manual verification, which is consistent with the results from prior studies~\cite{ROY2009470,NicadEvaluation}.

\phead{External validity.}
We conducted our study on five large-scale open source systems in different domains. We found that our uncovered patterns and the corresponding problematic and justifiable cases are common among the studied systems. However, our finding may not be generalizable to other systems. Hence, we studied whether the uncovered patterns exist in three other systems. We found that the patterns of duplicate logging code smells also exist in these systems and we did not find any new duplicate logging code smell patterns in our manual verification. Our studied systems are all implemented in Java, so the results may not be generalizable to systems in other programming languages. Future studies should validate the generalizability of our findings in systems in other programming languages.

%We only conducted our study on four studied systems, but our findings are widespread and generalizable in these four systems. We choose the studied systems with different sizes, across various domains in order to improve the generalizability. However, conducting the study on other more projects would further present the generalizability of our approach. We conduct the manual study and implement the duplicate logging code smells detecting tool, namely DLFinder, based on Java projects. Some of our findings may not able to be straightforwardly applied on projects based on other programming languages. However our approach of tool-assisted manual study is generic and generalizable, it can be used to uncover the duplicate logging code smells for other programming languages.

\vspace{-0.1cm}

%% file: texfiles/related.tex
\section{Related Work}
\label{sec:related}

\vspace{-0.1cm}
%\todo{Zhenhao, can you write the related work section? I feel writing this can help you get more familiar with the area and how this work is different from prior work. This can help you with defense :) You can check how Jack wrote their related section in the ICSE paper. In general, you need to categorize the related work into several areas, such as ``Detecting Logging Problems''. You also need to highlight the difference between our work and the work in each area. Usually you discuss the difference in the end of each area (in each subsection that discuss the work in each subareas)} \zhenhao{Sure, I'll work on it}
%In this section, we discuss three areas of related research: empirical studies on logging practices, improving logging practices, and studying and refactoring code smells.

\phead{Empirical studies on logging practices.}
%Previous studies show that software logs are extensively used and analyzed for various tasks, such as error diagnosis \cite{Yuan:2011:ISD:1950365.1950369, Yuan:2010:SED:1736020.1736038}, deployment verification \cite{Shang:2013:ADB:2486788.2486842}, load testing \cite{Chen:2017:ALT:3103112.3103144,jacktool}, understanding code quality \cite{Shang2015}, security monitoring\cite{DBLP:conf/dsn/MontanariHDBC12}, program comprehension \cite{Hassan:2008:ICS:1368088.1379445,Shang:2014:ULL:2705615.2706065}, and performance analysis\cite{Chen:2016:CHD:2950290.2950303,kundi_icpe_2018,DBLP:conf/nsdi/NagarajKN12}.
There are several studies on characterizing the logging practices in software systems~\cite{Yuan:2012:CLP:2337223.2337236, Chen2017,Fu:2014:DLE:2591062.2591175}. Yuan et al.~\cite{Yuan:2012:CLP:2337223.2337236} conducted a quantitative characteristics study on log messages for large-scale open source C/C++ systems. Chen et al.~\cite{Chen2017} replicated the study by Yuan et al.~\cite{Yuan:2012:CLP:2337223.2337236} on Java open-source projects. Both of their studies found that log message is crucial for system understanding and maintenance. Fu et al.~\cite{Fu:2014:DLE:2591062.2591175} studied where developers in Microsoft add logging statements in the code and summarized several typical logging strategies. They found that developers often add logs to check the returned value of a method. %Prior studies focus on studying where do people add logging code and how often do developers modify logging code. However, these studies do not analyze the quality of the log lines. In this study,
Different from prior studies, in this paper, we focus on manually understanding duplicate logging code smells. We also discuss potential approaches to detect and fix these code smells based on different contexts (i.e., surrounding code).

\phead{Improving logging practices.}
%\ian{add the new ASE papers}
%A number of studies have been conducted towards improving logging practices.
Zhao et al.~\cite{Zhao:2017:LFA:3132747.3132778} proposed a tool that determines how to optimally place logging statements given a performance overhead threshold. Zhu et al.~\cite{Zhu:2015:LLH:2818754.2818807} provided a tool for suggesting log placement using machine learning techniques. %, specifically in exception handling code blocks.
Yuan et al.~\cite{Yuan:2011:ISD:1950365.1950369} proposed an approach that can automatically insert additional variables into logging statements to enhance the error diagnostic information. Chen et al.~\cite{log_pattern_ICSE2017} concluded five categories of logging anti-patterns from code changes, and implemented a tool to detect the anti-patterns. Hassani et al.~\cite{mehran_emse_2018} identified seven root-causes of the log-related issues from log-related bug reports. Compared to prior studies, we study logging code smells that may be caused by duplicate logs, with a goal to help developers improve logging code. The logging problems that we uncovered in this study are not discovered by prior work. We conducted an extensive manual study through obtaining a deep understanding on not only the logging statements but also the surrounding code, whereas prior studies usually only look at the problems that are related to the logging statement itself. %, and encoded their findings into a static analysis tool that is able to detect the root-causes.

\phead{Code smells and code clones.}
%\peter{most of this need to be rewritten and need to be expanded significantly}
%\zhenhao{Might need to change the title if we mainly talk about code clones here}
Code smells can be indications of bad design and implementation choices, which may affect software systems' maintainability~\cite{7194592,Ahmed:2017:EER:3200492.3200502,6392174, dannydigmobisoft}, understandability~\cite{8115653, 6065171}, and performance~\cite{10.1007/978-3-319-26529-2_18}. To mitigate the impact of code smells, studies have been proposed to detect code smells~\cite{6693086,Nguyen:2012:DEC:2351676.2351724,Parnin:2008:CLV:1409720.1409733, Schumacher:2010:BES:1852786.1852797, DBLP:journals/ese/HermansPD15}. Duplicate code (or code clones) is a kind of code smells which may be caused by developers copying and pasting a piece of code from one place to another~\cite{5463343, tracyhallcodesmell}. Such code clones may indicate quality problems. %the log statements and their surrounding code to different places.
There are many studies that focus on studying the impact of code clones~\cite{DoCodeClonesMatter, kapser2006a,FrequencyAndRisksClones}, and detecting them~\cite{kamiya2002, cpminer, nicad}. In this paper, we study duplicate logging code smells, which are not studied in prior duplicate code studies. We also investigate the relationship between duplicate logging statements and code clones.
 Some instances of the problematic duplicate logging code smells in our study might also be related to micro-clones (i.e., cloned code snippets that are smaller than the minimum size of the regular clones~\cite{microclones}). A small number of prior studies investigate the characteristics and impact of micro-clones in evolving software systems~\cite{MicroclonesAndBugsSANER,MicroclonesAndBugsICPC,microclones,microclone4,microclone5}. Specifically, micro-clones may have similar tendencies of replicating severe bugs as regular clones~\cite{MicroclonesAndBugsICPC, MicroclonesAndBugsSANER}. However, the potential impact of micro-clones on logging code are not studied in these works. Our study provides insights for future studies on the relationship between micro-clones and logging code. The investigation on duplicate logging code smells and duplicate logging statements may also help identify micro-clones and further alleviate the impact of micro-clones on software maintenance and evolution.

%A number of studies also investigate the prevalence and characteristics of micro-clones (i.e., cloned code snippets that are smaller than the minimum size of the regular clones)~\cite{microclones, MicroclonesAndBugsSANER,MicroclonesAndBugsICPC}.

 \vspace{-0.1cm}

%We find that such duplicate logs may not necessarily be associated with duplicate code, and there are unique patterns of duplicate logging code smells that need to be refactored in order to help developers with debugging and log analysis. %still need specifical refactoring in order to improve the understandability of logs.

%Code clone detection techniques typically require setting a certain thresholds~\cite{roy09}. Non-optimal thresholds have significant impacts on the detected clones~\cite{roy09,7081830}, choosing the optimal thresholds is a non-trivial task and the value may differ across systems~\cite{nikos_icse_2018}. Duplicate logs are actually not necessarily caused by code clones but may still require refactoring. However to the best of our knowledge, there is no prior study which specifically tackles duplicate logs. Hence, comparing to the prior works which only address general code clone problems, we conduct our study specifically on duplicate logs and investigate the amount of duplicate logs which are associated with cloned code. We find that there is a significant number of duplicate logs that are not related to code clones but they still need a particular way to apply refactoring on them.

%% file: texfiles/conclusion.tex
\section{Conclusion}
\label{sec:conclusion}
Duplicate logging statements may affect developers' understanding of the system execution. In this paper, we study over 4K duplicate logging statements in five large-scale open source systems (Hadoop, CloudStack, Elasticsearch, Cassandra and Flink). We uncover five patterns of duplicate logging code smells. Further, we assess the impact of each uncovered code smell and find not all are problematic and need fixes. In particular, we find six justifiable cases where the uncovered patterns of duplicate logging code smells may not be problematic. We received confirmation from developers on both the problematic and justifiable cases.
%We reported both the problematic and justifiable cases to developers. All of the reported problematic instances are now fixed by developers, and developers agreed and acknowledged the uncovered justifiable cases.
Combining our manual analysis and developers' feedback, we developed a static analysis tool, \tool, which automatically detects problematic duplicate logging code smells.
We applied \toolS on the five manually studied systems and three additional systems. In total, we reported 91 problematic duplicate logging code smell instances in the eight studied systems to developers and all of them are fixed. \toolS successfully detects 81 out of the 91 instances.
We further investigate the relationship between duplicate logging statements and code clones, in order to provide a more comprehensive understanding of duplicate logging statements and duplicate logging code smells. We find that most of the problematic instances of duplicate logging code smells and almost half of the duplicate logging statements reside in cloned code snippets. Among them, a large portion reside in very short code blocks which might be difficult to detect using existing code clone detection tools.

Our study highlights the importance of the context of the logging code, i.e., the nature of logging code is highly associated with both the structure and the functionality of the surrounding code. Future studies should consider the code context when providing guidance to logging practices, more advanced logging libraries are needed to help developers improve logging practice and to avoid logging code smells. Our findings also provide an initial evidence on the prevalence of duplicate logging statements that reside in cloned code snippets, and the potential impact of code clones on logging practices. Future studies may also consider integrating different information in the software artifacts (e.g., duplicate logging statements) to further improve clone detection results.%, \zhenhao{Keep the following sentence or not?} and may highlight the potential of using duplicate logging statements to further improve clone detection tools.